\documentclass[%
 reprint,
 amsmath,amssymb,
 aps,superscriptaddress
]{revtex4-2}

\def\Rrr{\mathbf{R}_{\mathbf{rr}}}
\def\rout{\mathbf{r}_\textrm{out}}
\def\rin{\mathbf{r}_\textrm{in}}
\def\xout{x_\mathrm{out}}
\def\xin{x_\mathrm{in}}
\def\yout{y_\mathrm{out}}
\def\yin{y_\mathrm{in}}
\def\R{\mathbf{R}}

\usepackage[english]{babel}

\usepackage{graphicx}% Include figure files
\usepackage{dcolumn}% Align table columns on decimal point
\usepackage{bm}% bold mathS
\usepackage{hyperref}% add hypertext capabilities
\usepackage{xcolor}
\newcommand{\rita}[1]{\textcolor{black}{#1}}
\newcommand{\alex}[1]{\textcolor{black}{#1}}

\begin{document}

%\title{On the weight of single and recurrent back-scattering from random media}
\title{Distribution of seismic scatterers in the San Jacinto Fault Zone, southeast of Anza, California, based on passive matrix imaging}
% / Single and recurrent back-scattering from random media
\author{Rita Touma}
\affiliation{ISTerre, Universit\'{e} Grenoble Alpes, Maison des G\'{e}osciences, BP 53, F-38041 Grenoble}%
\affiliation{Institut Langevin, ESPCI Paris, CNRS, PSL University, 1 rue Jussieu, F-75005 Paris, France}%
\author{Alexandre Aubry}
\affiliation{Institut Langevin, ESPCI Paris, CNRS, PSL University, 1 rue Jussieu, F-75005 Paris, France}%
\author{Yehuda Ben-Zion}%
\affiliation{Department of Earth Sciences and Southern California Earthquake Center, University of Southern California, Los Angeles CA 90089, USA}% 
\author{Michel Campillo}
\affiliation{ISTerre, Universit\'{e} Grenoble Alpes, Maison des G\'{e}osciences, BP 53, F-38041 Grenoble}%

\date{\today}
\begin{abstract}
Fault zones are associated with multi-scale heterogeneities of rock properties. Large scale variations may be imaged with conventional seismic reflection methods that detect offsets in geological units, and tomographic techniques that provide average seismic velocities in resolved volumes. However, characterizing elementary localized inhomogeneities of fault zones, such as cracks and fractures, constitutes a challenge for conventional techniques.
Resolving these small-scale heterogeneities can provide detailed information for structural and mechanical models of fault zones.
Recently, the reflection matrix approach utilizing body wave reflections in ambient noise cross-correlations was extended with the introduction of aberration corrections to handle the actual lateral velocity variations in the fault zone~\citep{touma2021}.
Here this method is applied further to analyze the distribution of scatterers in the first few kilometers of the crust in the San Jacinto Fault Zone at the Sage Brush Flat (SGB) site, southeast of Anza, California. The matrix approach allows us to image not only specular reflectors but also to resolve the presence, location and reflectivity of scatterers for seismic waves starting with a simple homogeneous background velocity model of the medium. The derived three-dimensional image of the fault zone resolves lateral variations of scattering properties in the region within and around the surface fault traces, as well as differences between the Northwest (NW) and the Southeast (SE) parts of the study area. A localized intense damage zone at depth is observed in the SE section, suggesting that a geometrical complexity of the fault zone at depth induces ongoing generation of rock damage.
\end{abstract}

%\begin{description}
%\end{description}

\maketitle

\section{Introduction}
\label{Intro}
Earthquakes are among the most destructive natural disasters. Although earthquakes are generally unpredictable, some aspects of their behavior such as the likelihood of being arrested and statistically-preferred propagation direction can be estimated from structural properties of fault zones (see e.g.~\cite{chester1993,wesnousky1988,BZ2008}). Fault zones that are the structural manifestation of earthquakes evolve during deformation and have generally complex properties (e.g.~\cite{mitchell2009,BZandSammis2003}). Characterizing the geometrical and seismic properties of fault zones can provide important information for assessing likely past and future rupture properties. Fault zones have also strong impact on fluid flow in the lithosphere~\citep{knipe1998}. 

Fault zones are manifested at the surface by several main fault traces that accommodate the bulk of the long term slip. They are characterized by lineaments, topography and various geometrical complexities. The identification of fault traces is done by field observations (major line of fracturing, offset in geological units), along with remote sensing techniques that analyze ground deformation after major earthquakes obtained from satellites and aircrafts such as Interferometric Synthetic Aperture Radar (InSAR, ~\cite{massonnet1993}) and subpixel correlation of optical images (SPOT,~\cite{binet2005}). Fault zone properties below the surface are obtained by seismic and other geophysical imaging techniques.

Fault zones have hierarchical damage structures that evolve during the fault zone activity and have several general elements (e.g.~\cite{chester1993,rockwell2007}). The principle slip zone is a highly localized thin layer ($0.01-0.1$ m thick) that accommodates most of the fault slip and is characterized by ultra cataclasite rock particles. The principle slip zone is bounded by a core damage zone (inner damage zone) that is typically about 100 m wide and asymmetrically located on one side of the slip zone of large faults~\citep{lewis2005,dor2006}. The core damage is surrounded by a broader zone of reduced damage intensity (referred to as outer damage zone) that may extend for several km on each side of the fault. Properties of the fault zone damage provide information on statistical tendencies of local earthquake ruptures, operating dynamic stress field, energy dissipation and more (e.g.~\cite{manighetti2005,mitchell2009,xu2012}). For that reason, a number of seismic and other methods have been developed to provide detailed information on fault zone structures. 

Among the seismic imaging techniques, reflection seismology is generally pertinent to image planar horizontal layers with a very high resolution, and provides indirect imaging of faults by the offset of sedimentary layers. It relies on the analysis of seismic waves that are sent back towards the surface after being reflected or scattered by subsurface structures with strong impedance contrasts. The recorded wavefield is composed of reflected waves that result from the interaction of seismic waves with planar reflectors such as  layer boundaries, and diffracted waves from small-scale geological objects such as cracks  and fractures. To obtain structural information on the subsurface, migration techniques are applied aiming mainly to relocate reflectors and scatterers in depth or in time~\cite[see, e.g., review by][]{etgen2009}. Migration of seismic wavefields requires an accurate velocity model of the Earth. Errors and biases in the velocity model produce artefacts and defocusing due to phase distortions in the migrated images~\citep{moser2008}.

Reflection information is often gathered from seismic surveys where seismic energy propagating into the medium is generated by man-made sources such as vibrators, explosives, etc. In the last decade or so, passive methods based on the ambient seismic noise have been developed to substitute active imaging techniques~\cite[e.g.][]{campilloandroux2014}. Cross-correlation of passive traces recorded at two receivers allows retrieving the Green’s function between these two receivers. In other words, the resulting correlation is comparable to the seismogram that would be obtained at the first receiver if there is a source located at the second receiver’s location. This approach is referred to as seismic interferometry~\citep{wapenaar2010}. Noise cross-correlations are often used to retrieve the surface waves component of the Green’s function~\cite[][and later works]{shapiro2004}. Extracting body waves contributions to the Green’s function is more difficult~\citep{poli2012a}. However, it has been shown that ambient noise cross-correlations can be used to image deep targets inside the Earth with body wave reflections~\citep{draganov2007,poli2012b}. Inspired by works done in ultrasound imaging~\citep{aubry2009} and optical microscopy~\citep{badon2016}, a \say{reflection matrix approach} was introduced to geophysics and used body wave reflection from coda-wave cross-correlations to image the complex medium below Erebus volcano - Antarctica~\citep{blondel2018}.~\rita{The \say{reflection matrix approach} allows to image both specular reflectors, corresponding to the boundaries between layers of different propagation velocities, and non-specular reflections generated by a distribution of localized inhomogeneities.}
\rita{The matrix imaging approach} has then been developed to overcome phase distortions for multi-layered media~\citep{Lambert2020} and strongly heterogeneous media~\citep{Badon2019,lambert2021refle}. Using the same matrix formalism,~\cite{touma2021} analysed ambient noise recorded at a dense array \rita{($1108$ vertical geophones in 600 m $\times$ 700 m configuration)} to image subsurface properties of the San Jacinto Fault zone (SJFZ) southeast of Anza, California. While~\cite{blondel2018} dealt with imaging problems in the multiple scattering regime in the case of volcanoes, the main challenge in fault zones imaging is the presence of phase distortions, also referred to as aberrations, due to the strong structural heterogeneities within and around the fault zones.

\begin{figure*}
 \centering
 \includegraphics[width=14cm]{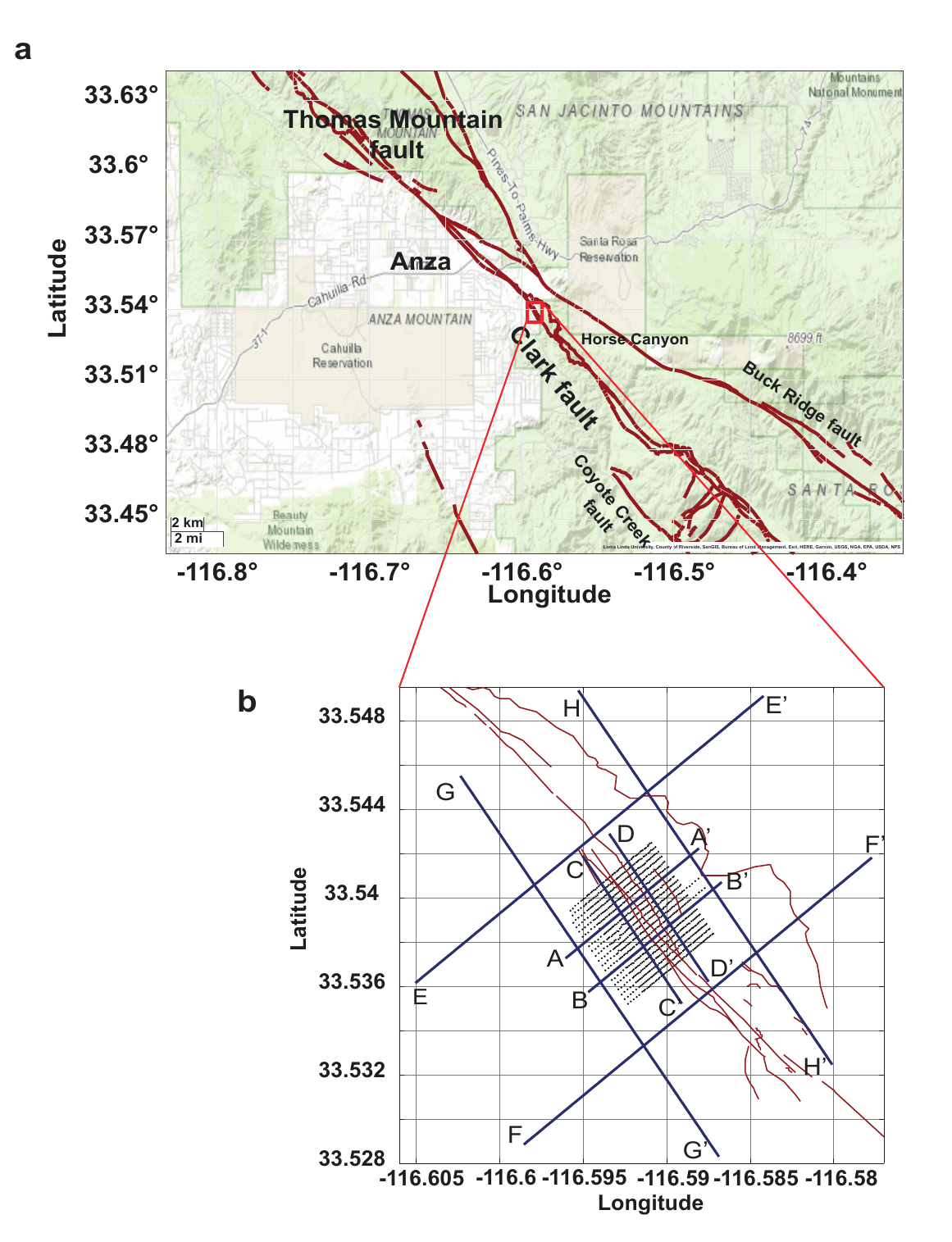}
 \caption{(a) Topographic map of Trifurcation area of San Jacinto fault zone. The red square marks the studied area. (b) Map of the geophones of the dense array at SGB site (black dots). The Clark fault traces are represented by the red lines. The blue lines indicate the locations of the cross-sections represented in Fig.~\ref{fig2} and~\ref{fig3}. }
\label{fig1}
\end{figure*}

The present paper follows the approach of~\cite{touma2021} to derive more detailed 3D images of the SJFZ at depth \rita{and to resolve the shallow structure with unprecedented resolution.}~\cite{touma2021} computed noise cross-correlations in the (10-20) Hz frequency range. Whereas the cross-correlations provide a response matrix between sources and receivers located at the surface, the reflection matrix contains the response between sources and receivers that are virtually moved inside the medium by performing focusing operations. This process is generally known as redatuming~\citep{Berkhout1993} and it allows local information on the medium’s reflectivity to  be retrieved. To project the data in the virtual focused basis, an homogeneous transmission matrix is used with a constant velocity model of $1500$ m/s. This velocity was chosen to optimize the focusing as discussed in~\cite{touma2021}.
The main advantage of this method is that it only requires an approximate estimation of the medium’s velocity. Although an incorrect velocity model produces phase distortions (aberrations) in the propagated data, the reflection matrix approach allows us to account and correct these aberrations through the distortion matrix concept~\citep{Badon2019,william,lambert2021distor}. From that matrix, the distorted component is extracted and is used to virtually focus back waves inside the medium (\rita{see \ref{method}}). As a result of the correction process, the resolution of the final subsurface images is drastically improved and a three-dimensional image of the subsurface reflectivity is revealed. The derived images represent reflectivity maps of the medium beneath the Clark branch of the SJFZ at the Sage Brush Flat site~\citep{Ben_Zion_2015,Roux_2016}. The site under study is located in the complex trifurcation area southeast of Anza, California (Fig.~\ref{fig1}a). The locations of the surface fault traces are derived from recent detailed studies of the surface geological mapping and shallow geophysical imaging~\citep{wade2018,share2020}. The basic goal of this paper is to interpret the obtained scattering images that exhibit features with higher lateral resolution than conventional seismic investigations. The variability and attenuation of scattered intensity within and around the major fault zone are also discussed.

\section{The San Jacinto Fault Zone}

The $230$ km-long San Jacinto Fault Zone is the most seismically active fault zone in southern California~\citep{hauksson2012} and is one of several major right-lateral strike-slip fault zones over which the North American-Pacific plate boundary is distributed in southern California. The SJFZ branches from the San Andreas fault at Cajon Pass and was formed $1-2$ million years ago, presumably in response to the geometrical complexities on the San Andreas Fault in the transverse ranges (e.g.~\cite{matti1993}). The SJFZ represents a less mature evolutionary stage in the life of a large continental strike-slip structure than the San Andreas fault. Approximately $24$ km of slip has been accommodated by the SJFZ since the latest Pliocene to early Pleistocene~\citep{dorsey2006}, with estimated slip rates that vary along strike between $8-20$ mm/yr~\citep{rockwell2003,fialko2006}. 
The SJFZ has varying surface complexity and seismicity along its strike. The Anza section to the northwest of the SGB site consists of a single strand, the Clark fault, with relatively regular geometry and low current background microseismicity~\citep{sanders1984}. The trifurcation area of the SJFZ where the SGB site is located (Fig.~\ref{fig1}) is associated with branching of the Clark fault in the Anza section into three major faults: a continuation of the Clark fault and the Buck Ridge and Coyote Creek faults. The Trifurcation area has a broad zone of high seismicity rates that include five earthquakes with magnitudes around 5 since 2001. The geometrical properties of the seismicity in the trifurcation area are very complex and consist of a diffuse pattern in the top $5$ km that changes to more localized structures dipping to the NE below $6$ km, along with zones of seismicity that are orthogonal to the main strike of the SJFZ~\citep{ross2017}.

\section{3D scattering volume}\label{images}

Migration techniques are known to be powerful tools for imaging strong reflecting boundaries. These  boundaries are identified by discontinuities of acoustic impedance in the subsurface and are characterized by specular returns in seismic records. Less interest has been accorded to the non-specular component that arises from small-scale geological objects~\citep{khaidukov2004}. The energy generated by such small objects is commonly referred to as diffractions. Non-specular energy holds valuable information on the local heterogeneities in the medium~\citep[and references therein]{schwarz2019}. \rita{While many conventional migration techniques, such as Gaussian beam migration and reverse-time migration, honor the diffracted component, such contributions are generally difficult to analyze: they are often suppressed due to the processing done in conventional seismic methods or masked by specular reflections whose amplitudes are much larger than the scattered components~\citep{kozlov2004}. They can also be considered as noise in several migration techniques.} Keeping the non-specular component in the analysis allows retrieving signatures of localized scatterers such as cracks or inclusions that lack lateral continuity. Imaging such features whose size is of the order and even smaller than the seismic wavelength contributes significantly to seismic interpretation~\citep{schwarz2020}.

Fault surfaces and zones with increased fractures density are non-specular objects for surface sensors~\citep{kanasewich1988}. Imaging such features is a challenge in most conventional seismic exploration surveys. Several studies have discussed the necessity of distinguishing between diffractions and reflections, and provided techniques to separate them~\citep{khaidukov2004,moser2008,bakhtiari2018}. Diffraction imaging is performed usually by suppressing the specular reflections so that the migrated image contains the diffracting component that have been isolated. The reflection matrix technique allows us to image, without performing any prior filtering, not only specular reflectors but mainly non-specular backscattered energy that directly gives insight into rock properties at the subsurface. The contribution of specular and non-specular features is distinguishable in the derived images. 
%The specular component is imaged partially due to the limited aperture size of the array~\rita{ (see supplementary Fig.~\ref{figS3})}.

The constant velocity model of $1500$ m/s chosen for the redatuming operation in the initial study of~\cite{touma2021} is highly approximate especially for the deep structure. It was chosen to optimize the focusing at depth with an "apparent" velocity that increases the effective aperture of the geophone array to exploit the actual contribution of multi-scattered paths as discussed in~\cite{touma2021}. We recall that a higher velocity will only stretch the detected features vertically. The images are presented as a function of depth for the chosen background velocity and two-way travel time.

\begin{figure*}
 \centering
 \includegraphics[width=14cm]{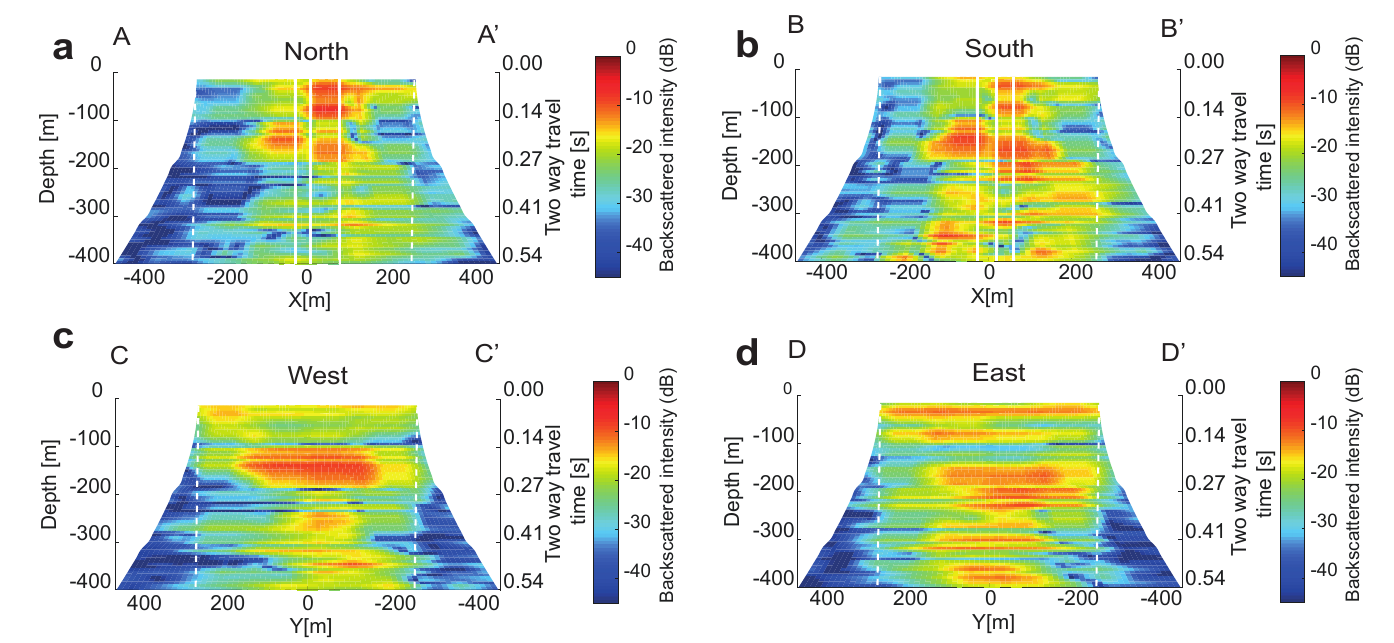}
 \caption{Shallow cross-sections of the 3D scattering volume. Vertical slices oriented perpendicular to the fault traces. North/South denote Northwest and Southeast. The main fault strands are represented by the bold white lines. (a) Profile A-A', (b) Profile B-B'. Vertical slices oriented parallel to the fault traces. West/East denote Southwest and Northeast. (c) Profile C-C', (d) Profile D-D'. The color scale is in dB. The white dashed lines correspond to the spatial extension of the array.}
\label{fig2}
\end{figure*}

\begin{figure*}
 \centering
 \includegraphics[width=14cm]{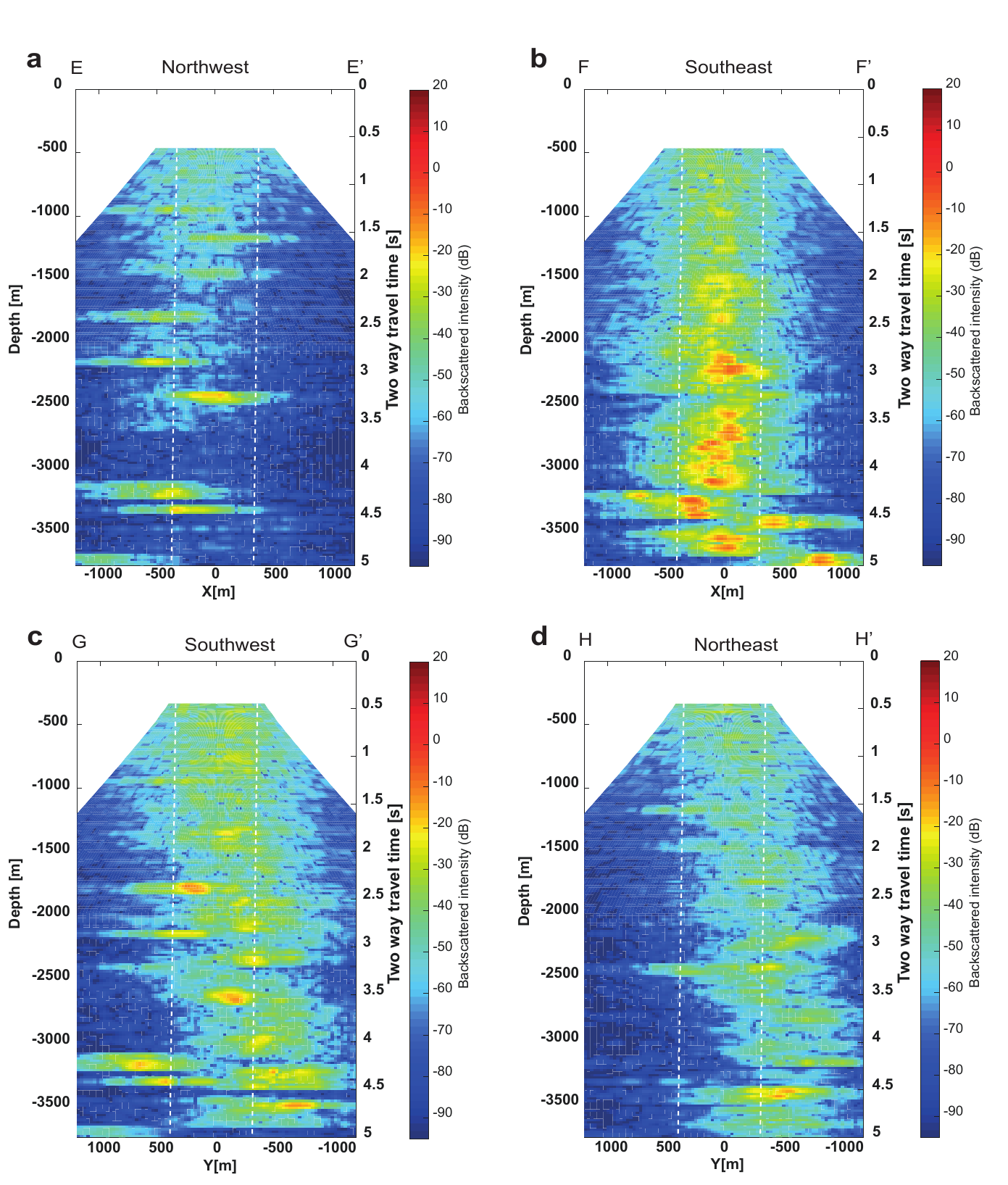}
 \caption{Deep cross-sections of the 3D scattering volume. Vertical slices oriented perpendicular to the fault traces. (a) Profile E-E' Northwest (NW) of the array, (b) Profile F-F' Southeast (SE) of the array. Vertical slices oriented parallel to the fault traces. (c) Profile G-G' Southwest (SW), (d) Profile H-H' Northeast (NE). The color scale is in dB. The white dashed lines correspond to the spatial extension of the array.}
\label{fig3}
\end{figure*}

In the subsequent subsections, we provide a detailed description of cross-sections taken from the 3D scattering volume. Those images are obtained after correcting the aberrations induced by the mismatch between the velocity model used to perform the focusing operation and the data. \alex{Although the images shown in Fig.~\ref{fig2} and Fig.~\ref{fig3} will lead to enlightening interpretations, they should be also taken with caution.} \rita{First, these images have been corrected from transverse aberrations but axial aberrations can subsist.}~\rita{To cope with this issue, a more accurate 2D velocity model can be implemented in the future.} %Another option is to consider a 3-D reflection matrix enabling in this case to account for axial aberrations and to relocate reflectors in depth~\citep{lambert2021distor}.} 
\alex{Secondly, the aberration correction process applied to these images only identifies two isoplanatic patches at each depth. This is probably a perfectible point in such complex heterogeneous media with strong lateral velocity variations. Other less reflective features belonging to other isoplanatic patches are thus possibly not revealed by our method in Fig.~\ref{fig2} and~\ref{fig3}.} 
%A solution to to that shorcoming will be to perform a local analysis of aberrations as recently shown by~\cite{lambert2021distor}.}

A compensation method for attenuation, described in ~\ref{compensation}, is applied to the reflectivity maps. 
The results are plotted on a Cartesian grid, with the origin located at the center of the array and the x-axis orthogonal to the fault traces. The colors represent the backscattered intensity plotted in dB. The local distribution of heterogeneities and discontinuities of material boundaries are revealed with maximal focused intensity.
We first present results associated with shallow materials and then discuss deeper structures.

\rita{We compared our results with previous studies investigating the same region. Most of the studies conducted at the SGB site utilize surface waves to resolve shallow features and the velocity distribution below the dense array~\citep{Roux_2016,hillers2016, mordret2019} or fault zone head waves to image lateral variations of lithology and damage across the fault~\citep{Ben_Zion_2015,qin2018}. To our knowledge, the deep scattering structure has not been examined and analyzed yet.}

\subsection{Images of the shallow fault zone}

The high frequency cross-correlations of noise recorded by the dense array allow resolving features in the top few hundred meters of the crust. Fig.~\ref{fig2} shows vertical slices of the 3D volume with a close-up view of the first $400$ m: two slices perpendicular to the fault (AA’, BB’) and two slices parallel to the fault traces (CC’, DD’). For the sake of simplicity, the cross-sections are labeled North, South, West and East, respectively. The slices are plotted in logarithmic dB scale and reveal the backscattered intensity rising from highly reflective features detected through the aberration correction process. The location of the each cross-section is indicated by blue lines in Fig.~\ref{fig1}b. In the first two vertical slices, the white lines refer to the location of the three main sub-parallel strands represented in red in Fig.~\ref{fig1}b.

Several differences between the results in the slices oriented differently stand up. The first thing to notice is the structural variations across the fault zone in Figs.~\ref{fig2}a and b.
We observe high intensity of scatterers within the core fault damage zone and reduced scatterers intensity outside. A clear offset of reflective structures is observed around a depth of $150$ m in the two perpendicular panels. Figs.~\ref{fig2}a and b share the same features although the scattering appears stronger and more extended in the southern profile (B-B’). The intensity below $150$ m decreases in the North cross-section while in the South cross-section a high density of scatterers extents to greater depth revealing a localized damage zone around the fault traces (white lines). 

The offset of the scatterers in Figs.~\ref{fig2}a and b is observed mainly below the SW fault trace.~\cite{qin2018} suggested that the SW fault trace is the main seismogenic fault separating two crustal blocks of different seismic properties.~\cite{mordret2019} also showed the presence of a velocity contrast across the SW fault trace. The observed offset in structural properties  can be explained by the fact that the SW trace represents the main seismogenic fault.

Figs.~\ref{fig2}c and d representing the West and East cross-sections show no clear lateral variations of the subsurface structures. The reflectivity is associated with planar features or layers located on each side of the fault. The high scattering zone to the NE of the surface trace, clearly observed in Fig~\ref{fig2}d, is in general agreement with the trapping structure identified by~\cite{Ben_Zion_2015} and~\cite{qin2018}. This zone is characterized by significant low velocities and an intense localized damage producing reflections~\citep{Roux_2016,hillers2016,mordret2019}. The reflective layer SW of the fault observed in Fig~\ref{fig2}c coincides with the local sedimentary basin reported by~\cite{Ben_Zion_2015},~\cite{Roux_2016} and~\cite{hillers2016}.

Many studies of the San Jacinto Fault Zone observed an asymmetric rock damage across the fault~\citep{lewis2005,dor2006,qin2018,wade2018}. The damage at the SGB site was shown to be greater on the NE side of the fault. The scattering in Fig.~\ref{fig2}a is more pronounced in the NE. Between $150$ and $400$ m, the scattering zone  dips slightly to the NE and is comparable to the shape of the low velocity trough found by~\cite{mordret2019} beneath the fault trace at the same location as profile AA’.

\subsection{Deeper and larger scale structure}\label{deepstruct}

We now investigate deeper sections from the 3D volume. Fig.~\ref{fig3} shows four vertical $4$ km deep slices; two slices are oriented perpendicular to the fault, one in the Northwest (EE’) of the array and the other in the Southeast (FF’), and two additional slices oriented parallel to the fault, one in the Southwest (GG’) of the array and the other in the Northeast (HH’). The locations of the cross-sections are marked by blue lines in Fig.~\ref{fig1}b.
\rita{All vertical cross-sections are presented in the form of animated movies in Animation D.1 (slices perpendicular to the fault traces) and Animation .D.2 (slices parallel to the fault traces).}

~\rita{In general, reflectors can only be imaged over a transverse field-of-view limited by the size of the geophones' array. Indeed, in the case of planar interfaces, a part of the reflected wave-field is not captured by the geophone array for large angles of incidence (See supplementary Fig.~\ref{figS3}a). Yet, the 3D scattering images shown in Fig.~\ref{fig3} reveal features that extend well beyond the spatial extension of the array. This observation is made possible due to the intense damage around the fault zone. In such diffusive media, the strong heterogeneities reflect the incident waves in many different directions, and therefore the imaged field-of-view is extended (See supplementary Fig.~\ref{figS3}b). Interfaces between geological layers are also imaged beyond the array extension by means of the waves scattered by the localized heterogeneities that reside at the discontinuity across interfaces.} 
%It is important to remember that one cannot deny that a better focusing and a higher intensity is obtained below the array's dimensions than on the edges of the array, especially in the case of a limited aperture array. However, the gain in the focusing quality assisted by the correction process enables to obtain a high resolution for structures imaged outside the array's dimensions.}

The comparison between the first two panels (Figs.~\ref{fig3}a and b) reveals a clear difference in reflectivity between the Northwest (NW) and Southeast (SE) portions of the SGB site. Both panels show a broad scattering zone in the shallow crust that has a V shape with about $800$ m wide area at $z=500$ m ($t=0.7$ s) (Figs.~\ref{fig3}a and b). 
%~\ritaremove{ decreasing width and intensity with depth. This zone has} 
The dense distribution of scatterers in the shallow zone results likely from the heavily damaged rocks around the fault traces. 
%~\ritaremove{, and it confirms that the damage intensity is rapidly decaying with depth in the first few kilometers of the crust~\citep{bZandshi2005,finzi2009,kaneko2011}}
In the NW panel, the diffuse damage is less apparent deeper than $z=1000$ m ($t=1.3$ s). However, the observed back-scattered energy in that section is associated with horizontal reflectors emerging on both sides of the fault. Discontinuous blocks on the right and left side of the fault traces highlight the offset of geological features across the fault. In contrast, the high intensity scattering zone extends deeper than $z=1000$ m ($t=1.3$ s) in the SE slice. Around $z=1500$ m ($t=2$ s), the backscattered intensity reveals a zone that is about $450$ m wide. Deeper in the crust, scattering seems to persist in combination with specular reflections arising from discontinuous deep layers ($z=3300$ m, $t=4.4$ s).

The different scattering zone extensions in Figs.~\ref{fig3}a and b are consistent with a change in the nature and structural complexity of the fault zone in the study area. To the NW of the SGB site, the SJFZ occupies a linear valley, whereas, to the SE, it becomes more localized and is associated with a canyon~\citep{sharp1967}. Recent geological mapping in the area (Wade 2018) shows multiple fault strands at the SGB site (Fig.~\ref{fig1}b). One main fault is mapped at the base of the NW boundary of the SGB basin. To the SE along strike, that fault merges with two other faults and results in a more localized zone that is associated with a higher damage intensity. The reflectivity panels confirm this feature by showing a significant scattering at depth SE of the array related to highly damaged (cracked and crushed) rocks. In the first $2$ km ($t=2.7$ s), the fault appears to be more localized in the SE generating an intense distribution of scatterers. Indeed, the scattered wavefield dominates and the specular component is less apparent in the SE compared with the NW. In the NW, the damage zone is more distributed and less intense. The damage intensity is rapidly decaying with depth, and the spreading of the scattering zone is mainly observed in the first kilometer. At larger depth, specular reflections predominate over the scattered component.

The panels oriented parallel to the fault also display a structural difference between the right and left sides of the fault. The scattering appears to be more concentrated in the SW profile (Fig.~\ref{fig3}c), whereas the NE profile (Fig.~\ref{fig3}d) shows strong continuity of planar boundaries. These observations are also in agreement with \cite[][Fig.3]{sharp1967} where the SW of the SJFZ at Table Mountain exhibits more complexity than the NE region. The scattering zones in both sections are dipping toward the SE where the fault zone is more localized. 

\rita{In the following, we will show that the reflection matrix contains much more information than the medium's reflectivity. It also provides a direct insight into the variations of the scattering properties and the attenuation of scattered waves in the fault region.}

%Also, the first two eigenstates considered in the correction process tend to correct over the two main isoplanatic patches which result on focusing on the more reflective structures. However, limiting the field of view to only two isoplanatic patches remains a hypothesis in such complex heterogeneous media with lateral velocity variations. A solution to compensate for phase distortions over several isoplanatic patches is to perform a local correction by dividing the field-of-view into a number of over-lapping regions~\cite{lambert2021distor}. This approach allows to resolve features located at the edges of the field-of-view and therefore improves the resolution in those regions. The number of resolution cells mapping each sub-region should be enough to be able to extract the spatial correlations of the distorted wave-fronts that converge towards the focusing law of the considered region. Because of the limited aperture array, and consequently the limited number of resolution cells, this process is not favored in the present case study and will be addressed in future works.}

\section{Lateral variations of intensity}

Seismic waves propagating inside the Earth give direct insight on the nature and properties of rocks. While travelling through complex heterogeneous media, waves suffer from seismic attenuation. In this section we briefly discuss seismic wave attenuation principles and how attenuation is accounted for in our matrix formalism, and therefore in the images obtained.

Seismic attenuation describes the decay of energy experienced by seismic waves while they propagate. Amplitudes are easily altered by several factors such as geometrical spreading, scattering and absorption (intrinsic or anelastic attenuation)~\citep{shapiro1993,knipe1998}. Evaluation of the attenuation due to scattering and intrinsic absorption has been the subject of considerable studies in seismology~\citep{aki1969,sato2012}. Estimating attenuation properties provides complementary information to seismic velocity distribution, and can    be particularly useful in fault zone studies to obtain a better understanding of rock properties and subsurface structures. Within the framework of the present paper, seismic attenuation is being examined to: (i) compensate the intensity decay in the 3D volume (see~\ref{compensation}) and (ii) reveal lateral variability of backscattered intensity in the fault zone.

\begin{figure*}
 \centering
 \includegraphics[width=14cm]{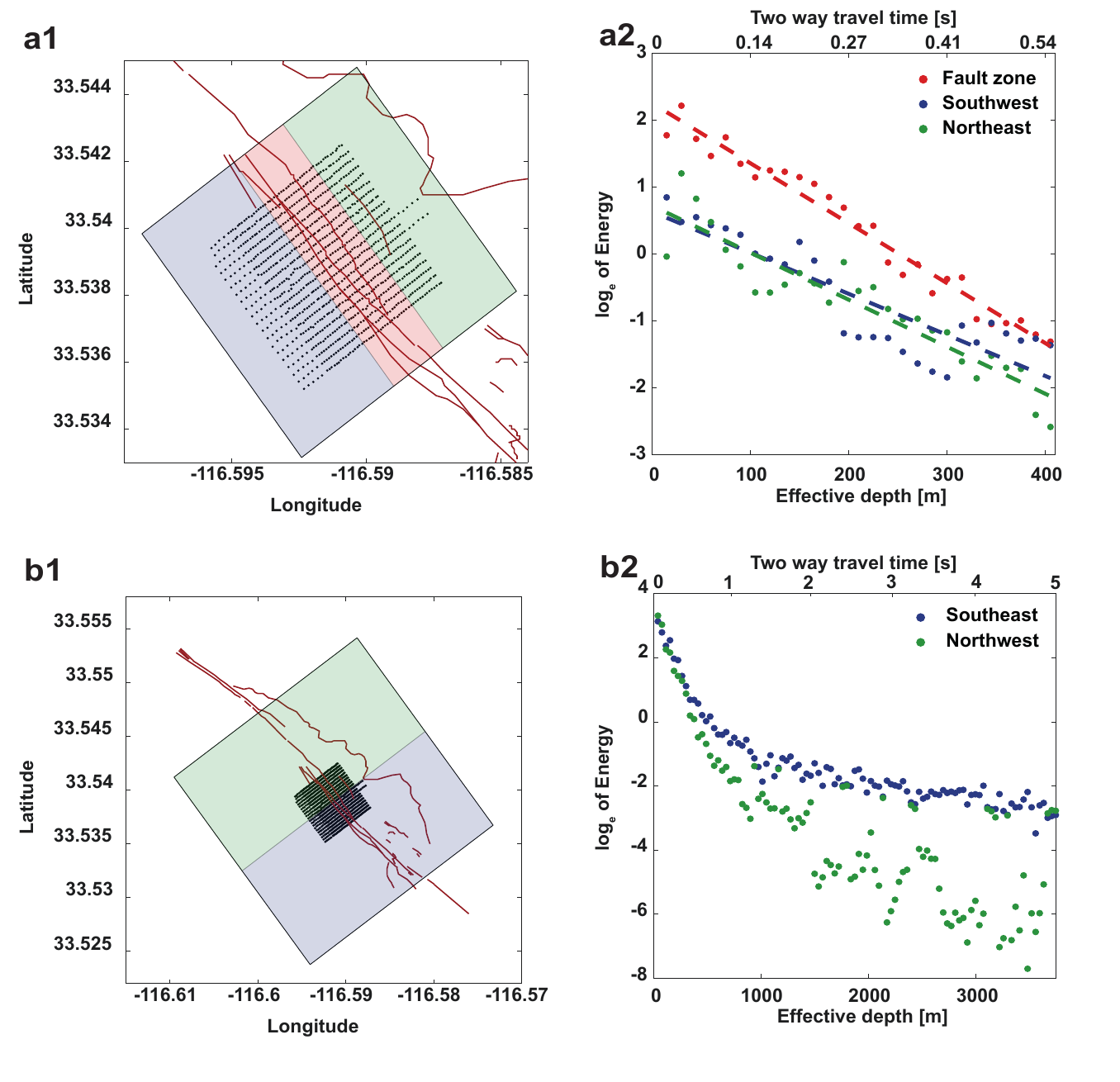}
 \caption{Time decay of the backscattered intensity. (a) Lateral variation of the time decay in the shallow crust for the first $400$ m. The area is discretized into three zones: red color refer to the fault zone, blue color to the SW and green color to the NE.
(b) NW - SE variation of the backscattered intensity time decay in the deep crust up to $4000$ m. The area is discretized in two zones: blue color refers to the SE and green color to the NW.
(1) Map of the geophones of the dense array at SGB site (black dots). The Clark fault traces are represented by the red lines. The black rectangle corresponds to the area covered by the images.
(2) Intensity decay as a function of apparent depth and, equivalently, of two way travel time of the scattering volume.}
\label{fig4}
\end{figure*}

The intensity represented in the pixels gives a direct estimate of the scattering properties in the region. To detect possible lateral variability of the energy distribution below the dense array, we divide  the area into sub-regions and examine the intensity of temporal decay for shallow and deep parts. In Fig.~\ref{fig4}a, the study area is divided into three zones displayed in Fig.~\ref{fig4}a1: a zone representing the main fault zone (red shaded area), a region to the SW of the fault zone (blue shaded area) and a region to the NE of the fault zone (green shaded area). For each region, we compute the mean intensity of the pixels located beneath that region. The energy distribution for the first $400$ m is presented in Fig.~\ref{fig4}a2. The linear regression is also plotted (dashed lines) to visualize the slope of the energy decay. We only show the energy decay corresponding to the images after the correction process. 
The first noticeable feature is that the fault zone is associated with higher intensity values resulting from non-specular energy transmitted from the localized damage. This is consistent with the overall amplification of seismic waves in low velocity fault zone layers (e.g.~\cite{kurzon2014}).
The energy in the fault zone with a steep slope distinguishes itself from the surrounding western and eastern regions where the decay slopes are more gradual. The fault zone is defined by major fracturing and crushed rocks, and consequently it is characterized by a rapid energy decay. In the neighboring regions where the damage is more distributed, less attenuation is observed in comparison with the fault zone. These results are consistent with significantly lower values of attenuation coefficients generally found within fault zones by waveform fitting of trapped waves (e.g.~\cite{lewis2005,qin2018}).

We also compare the intensity decay between the SE and NW sections $4$ km below the array. Fig.~\ref{fig4}b2 shows plots of the backscattered intensity averaged across the two regions delineated in Fig.~\ref{fig4}b1. We choose to plot the measured intensity values without the linear regression. The discrepancy observed in the backscattered intensity distribution primarily reflects the difference in subsurface structure between the two regions that was highlighted in section~\ref{deepstruct}. The fluctuation of intensity in the NW plot (green dots) is associated with the specular returns at several depths. In other words, high values correspond to the reflective boundaries observed in Fig.~\ref{fig3}a, while  the blue scatter plot representing the SE of the array decreases smoothly. This is explained by the consistent density of scatterers around the core of the fault zone damage area previously highlighted by the cross section in Fig.~\ref{fig3}b.

\section{Discussion and conclusions}

The presented results provide detailed images of seismic properties in the 3D volume around the San Jacinto fault at the SGB site. We used one month of ambient seismic noise recorded by a dense array deployed at SGB site around the Clark branch of SJFZ. The high frequency seismic data provided by the spatially dense array allows us to resolve features near the surface with high resolution. The reflectivity maps representing slices of the 3D scattering volume are obtained through the reflection matrix procedure developed in~\cite{touma2021}. Body wave reflections from ambient noise correlations are used to image the fault zone structure up to $4$ km below the surface. These images reveal the backscattered intensity generated by the distribution of heterogeneities in the medium.

Fault zones are very complex regions with extensive fracturing and damage that can reach the bottom of the seismogenic zone in some places as seen in tomographic studies around large faults (e.g.~\cite{allam2014}). Tomographic and other imaging studies provide average properties of rock volumes, but do not resolve the presence, location and intensity of scatterers that are imaged with the reflection matrix method. The strong variations of velocities and significant attenuation in fault zone regions present challenges for conventional imaging techniques. However, our reflection matrix approach allows us to derive the distribution of scatterers inside the medium with an approximate velocity model of the medium. To that aim, a powerful aberration correction process is performed and provides high resolution images of the subsurface. However, associating a reflector with a specific depth remains dependent on the reference velocity model, so the reflectivity maps are displayed as a function of an effective depth and the observed two-way travel time.

A significant advantage of the matrix approach is that focusing inside the medium enables the imaging of not only specular reflectors but also of scattering objects such as cracks and fractures. While many methods consider the diffracting and scattering components as noise in the seismic data, and tend to remove these components to image discontinuous layers, the current approach takes advantage of the scattering in the complex fault zone to resolve features of the order of the wavelength. This constitutes one of the main strength of the method. The images in Figs.~\ref{fig2} and~\ref{fig3} show both discontinuities of some layers that are signature of a large fault, along with lateral and axial variations in the backscattered intensity induced by cracks and other small-scale heterogeneities. The axial variations of the backscattered intensity in Fig.~\ref{fig4} also reveal systematic differences in scattering properties in the region within and around the surface fault traces relative to the outside volume, as well as differences between the NW and SE portions of the study area. The results are consistent with more localized intense damage zone at depth in the SE section where the SJFZ enters the Horse canyon, and more diffuse rock damage to the NW where the SJFZ is in a linear valley. The higher damage at depth in the SE section also suggests a geometrical complexity at depth leading to an ongoing generation of rock damage that is overprinted on older healed damage~\citep{sharp1967}. 

\begin{figure*}
 \centering
 \includegraphics[width=14cm]{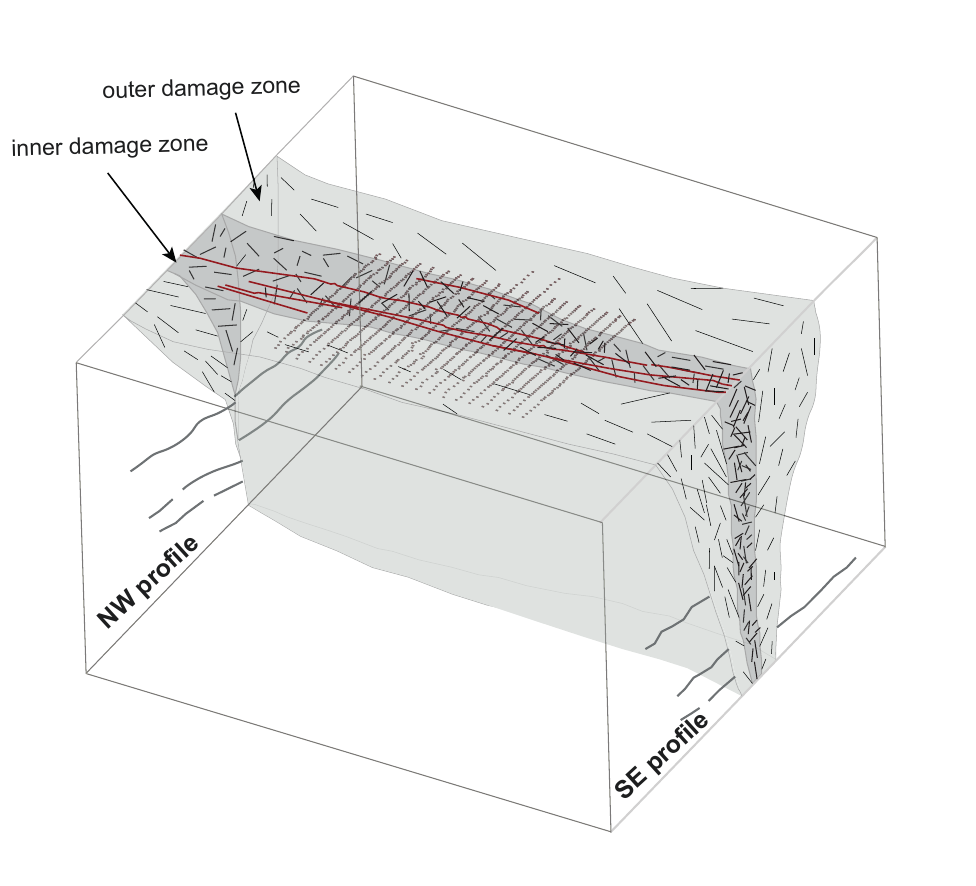}
 \caption{Schematic illustration of the fault zone at SGB site. Black dots refer to the geophones and red lines represent the Clark fault traces.}
\label{fig5}
\end{figure*}

Fig.~\ref{fig5} summarized schematically the obtained imaging results for both the inner and outer damage zones in the area. The fault traces in Fig.~\ref{fig1}a suggest a broader and less intense fault zone in the NW than in the SE. Surface observations are consistent with the fact that the main fault and the surrounding core damage are more localized in the SE where the principle slip zone is delineated by intense damage and fracturing extending down to $3$ to $4$ kilometers. In the NW, the scattering in the FZ is only observed in the first kilometer indicating a shallow less intense and diffuse damage. The results in Figs.~\ref{fig3}c and d show that the damage distribution is more complex to the SW side of the fault exhibiting a more pronounced outer damage in the SW than in the NE side of the fault.

The lateral variations of the fault structure between the NW and SE are consistent with the transition from the Anza section of the SJFZ associated with a single major fault trace, to a complex fault zone in the trifurcation area with several traces at the surface. The high scattering zone that extended to depth in the SE part of the SGB site is likely associated with highly damaged fault zone rocks between the sub-parallel strands in the SE.

\section*{Acknowledgments}

%\acknowledgments
We acknowledge support from the European Research Council (ERC) under the European Union Horizon 2020 research and innovation program (grant agreement No 742335, F-IMAGE and grant agreement No. 819261, REMINISCENCE). YBZ acknowledges support from the Department of Energy (Award DE‐SC0016520). 
%The study was funded primarily by LABEX WIFI (Laboratory of Excellence within the French Program Investments for the Future, ANR-10-LABX-24 and ANR-10-IDEX-0001-02 and by TOTAL R\&D.
The paper benefited from constructive comments by two anonymous referees and Editor Rebecca Bendick.

\section*{DATA AVAILABILITY}
The seismic data used in this study can be obtained from the~\cite{vernon2014} dataset in the Data Management Center of the Incorporated Research Institutions for Seismology (IRIS). The facilities of IRISData Services, and specifically the IRIS Data Management Center, were used for access to waveforms, related metadata, and/or derived products used in this study. IRIS Data Services are funded through the Seismological Facilities for the Advancement of Geoscience (SAGE) Award of the National Science Foundation under Cooperative Support Agreement EAR-1851048.

\appendix
\renewcommand{\thefigure}{S\arabic{figure}}
\renewcommand{\theequation}{S\arabic{equation}}

\section{Imaging and aberration correction} 
\label{method}

\setcounter{figure}{0}    

\begin{figure*}
 \centering
 \includegraphics[width=14cm]{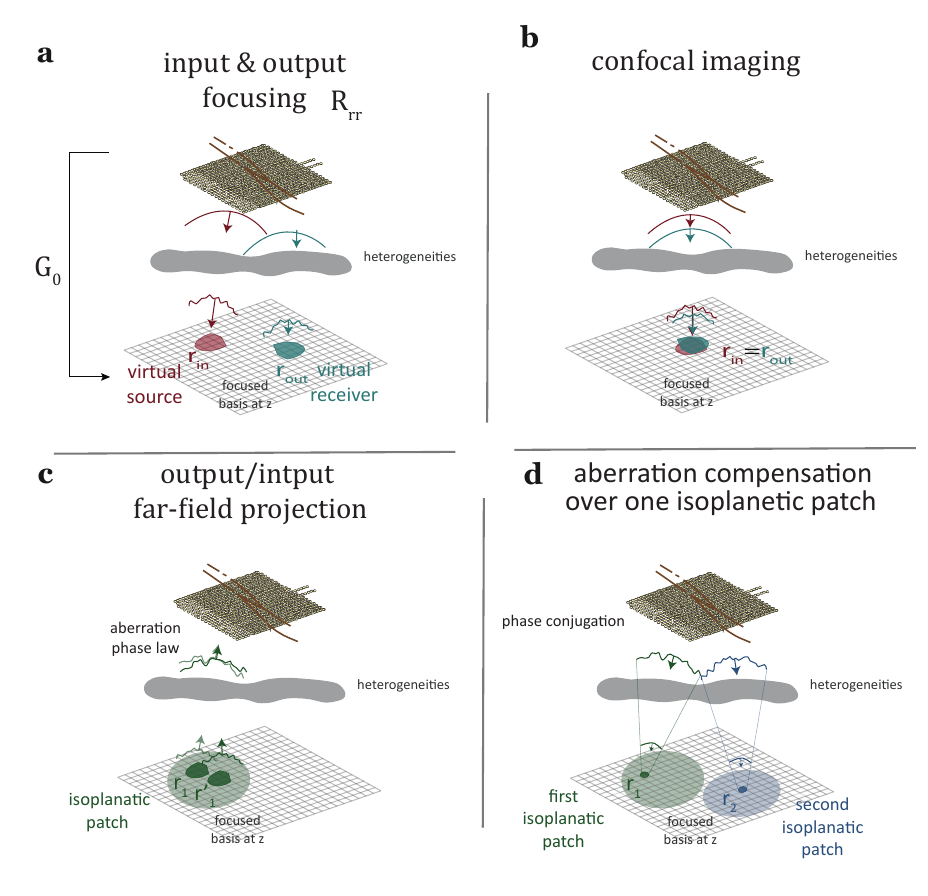}
 \caption {\rita{Focused reflection matrix and aberration correction. (a) The response matrix is projected onto a focused basis at depth $z$ both at emission (in) and reception (out). The focusing points synthesize virtual sources and receivers scanning every point in the virtual plane. In the presence of heterogeneities, the waves are distorted and the focusing is not optimal. (b) Confocal imaging principle: The image of the medium is obtained when input and output focusing are performed at the same location in the focused basis. (c) Far-field projection of the focused reflection matrix: For each virtual source, the reflected wave-front is investigated in the far-field. The phase distortions are identical for nearby virtual sources belonging to the same isoplanatic patch. (d) The phase conjugate of the aberration phase law enables a fine compensation of aberrations over the corresponding isoplanatic patch.}}
\label{figS1}
\end{figure*}

~\rita{In this section, we review the reflection matrix technique~\citep{touma2021} applied to obtain the 3D scattering images of the accompanying paper. First we describe the virtual focusing method inside the medium and then we recall the aberration correction process.}

~\rita{Ambient noise cross-correlations between the stations are computed in the 10-20 Hz range. These correlations provide the impulse response between virtual sources (emission) and receivers (reception) at the surface. In other words, the cross-correlations form a response matrix $\mathbf{K}(t)$ of the underground associated with the geophones' array. This matrix contains considerable of information about the medium, especially the distribution of its local reflectivity. This information is retrieved from the response matrix by performing focusing operations, both in emission and reception. Focusing consists in applying appropriate time delays, at emission and reception, so that the seismic waves resulting from a scattering event at the focal point interfere constructively. The goal of this operation is to virtually move the sources and receivers inside the medium onto a virtual plane. It is commonly referred to as redatuming~\citep{Berkhout1993} and can be easily implemented in the frequency domain via simple matrix products.
The first step is to apply a temporal Fourier transform to the correlation matrix to obtain  $\mathbf{K}(f)$. }

~\rita{To perform focusing, we define at each depth a basis of focal points $\mathbf{r}$, which corresponds to the location of virtual geophones inside the medium. A constant velocity model of $1500$ m/s is chosen for the transmission matrix $\mathbf{G_0}(f)$ that describes the propagation of waves between the surface and the focused basis. At each frequency, the response matrix is projected into the focused basis both in emission and reception (Fig.~\ref{figS1}a). The broadband focused reflection matrix $\Rrr(z)$ is then computed by calculating the coherent sum of the focused reflection matrices over the considered frequency range. This last operation amounts to a ballistic time gating of singly-scattered waves at time $t \sim 2z/c_0$. 
The coefficients $\R$ $(\rout,\rin)$ of this matrix correspond to the responses between a set of virtual sources $\rin=(\xin,\yin,z)$ and receivers $\rout=(\xout,\yout,z)$ at each depth $z$.
Among all these coefficients, the diagonal elements {($\rin=\rout$)}, are of particular interest since they provide a confocal image of the underground (Fig.~\ref{figS1}b). The other coefficients of $\Rrr$ are also useful since the focusing quality can be directly quantified by the spreading of the backscattered energy over its off-diagonal elements. In the presence of strong phase distortions (aberrations), resulting from a very complex seismic velocity distribution, each input and output focal spot can actually spread well beyond the diffraction limit, giving rise to a loss of resolution and contrast of the confocal image. Interestingly, the reflection matrix can also be used to retrieve in post-processing the shape of wave-fronts that would allow a perfect compensation for these phase distortions.}~\alex{To do so, one can exploit the spatial correlations that exist between the phase distortions urdergone by the reflected wave-fronts induced by nearby virtual geophones. This coherence area is referred to the isoplanatic patch (Fig.~\ref{figS1}c). The idea is to project the focused reflection matrix at input or output in a basis that maximizes the size of such isoplanatic patches in the focal plane. For a multi-layered basis, the most adequate basis is the Fourier basis ($k$) that amounts to projecting the reflected wave-field in the far-field.  The aforementioned spatial correlations are then leveraged by extracting the distorted component of each reflected wave-field.  It results in a distortion matrix whose singular value decomposition provides a decomposition of the field-of-view into a set of isoplanatic patches with the associated aberration phase transmittances. The phase conjugate of each aberration phase law provides the focusing law that should be used to compensate for aberrations on each isoplanatic patch (Fig.~\ref{figS1}d). Such a focusing law is applied here to the reflection matrix from the dual basis to obtain an updated focused reflection matrix. A corrected confocal image is obtained by considering the diagonal coefficients of this new focused reflection matrix.}
%~\rita{However, it is possible to quantify and correct for the phase distortions by extracting the phase law that can be used to compensation for these phase distortions. The correction allows to focus back waves inside the medium. }

~\alex{The whole process can be iterated at input and output in order to refine our estimation of the focusing law over each isoplanatic patch.}
\rita{To that aim, the focused reflection matrix should be projected in the far-field (Fourier basis $k$) by switching, at each iteration, between input or output.}

\section{Specular vs. diffuse reflection} \label{specularanddiffuse}

\begin{figure*}
 \centering
 \includegraphics[width=14cm]{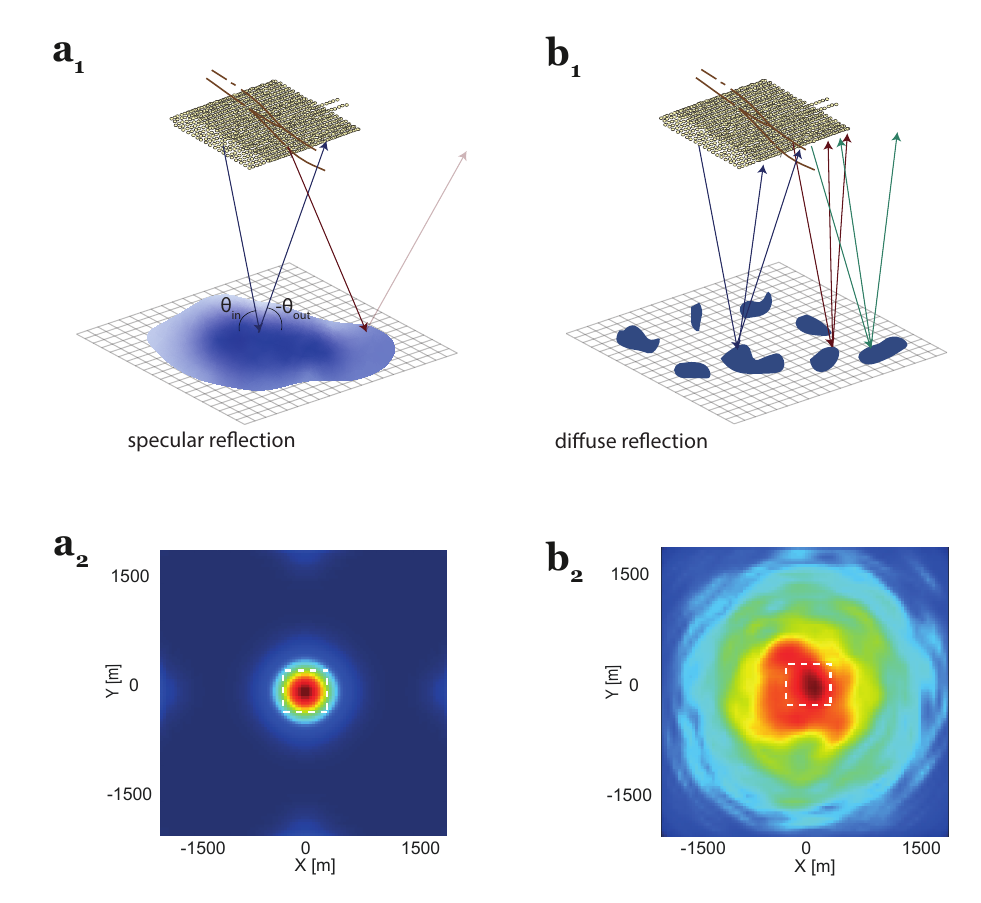}
 \caption { Impact of specular and diffuse reflection on the field-of-view of the confocal image. (a) Smooth mirror-like surface. (a.1) Sketch showing the reflected waves traveling with the same angle as the incident waves. Some of the reflected rays are not captured by the array. (a.2) Synthetic confocal image of a specular reflector at $z = 3600$ m. The white dashed square corresponds to the spatial extension of the array. (b) Random scattering medium. (b.1) Sketch displaying distributed heterogeneities at depth.
An incident ray is scattered in all directions enabling a confocal image that extends well beyond the lateral dimensions of the geophone array. (b.2) Mean synthetic confocal image at $z = 3600$ m estimated from several iterations obtained in the case of a random medium. The white dashed square corresponds to the spatial extension of the array. The imaged field-of view extends well beyond the transverse size of the array. }
\label{figS3}
\end{figure*}

\alex{The finite size of the geophone array has an impact on the field-of-view that can be imaged by the reflection matrix approach. To quantify this effect, two synthetic tests have been performed: (i) considering planar reflectors at a given depth $z = 3600 m$ (specular scattering, see Fig.~\ref{figS3}a1); (ii) considering randomly distributed scatterers at the same depth (diffuse scattering, see Fig.~\ref{figS3}b1). In each case, an homogeneous effective wave velocity $c_0 = 1500 m/s $ is considered. Fig.~\ref{figS3}a2 and~\ref{figS3}b2 show the results of the numerical simulations.}

\rita{Fig.~\ref{figS3}a2 shows a synthetic confocal image obtained for the specular reflector. The field-of-view is limited by the size of the array. The fraction of the incident wave-front that spreads beyond the transverse size of the array gives rise to reflected waves that cannot be recorded by the array of geophones (red arrows in Fig.~\ref{figS3}a1). Fig.~\ref{figS3}b2 displays the field-of-view of our matrix imaging method in the diffuse scattering regime. This intensity distribution is obtained by averaging the synthetic confocal image over a number $N=50$ realizations of disorder. Unlike for specular reflectors, incident wave-fronts are reflected in different directions (arrows in Fig.~\ref{figS3}b1). Consequently, the field-of-view spreads well beyond the transverse size of the geophone's array. The latter observation explains the wide field-of-view obtained in the 3D-scattering images of the accompanying paper (Fig.~\ref{fig3}). The reflection matrix method mainly images a distribution of small heterogeneities that are located in the damage area at shallow depths ($z < 1500 m$) and that reside between each geological layer at larger depths. }

%In both cases, the medium is supposed homogeneous with a velocity of $1500$ m/s. It is clearly noticeable that the planar reflector can only be imaged below the array, whereas in scattering media, the backscattered intensity is measured far outside of the array

\section{Time-gain compensation} \label{compensation}

The energy losses increase with the heterogeneity of the medium. The propagation matrix used to project the raw data to the focused basis accounts for geometrical spreading whereas the effect of
absorption and scattering are ignored. In particular, in strong scattering regime, these losses can strongly degrade the contrast of the images at larger depth. To overcome these problems in the imaging process and in order to visualize the entire field-of-view, it is mandatory to compensate for the amplitude drop in the 3D-images (Fig.~\ref{fig2} and~\ref{fig3}), especially in the shallow crust. This time gain compensation is done manually by multiplying the intensity profiles by an increasing function with depth, that is in this case the reciprocal of the mean intensity calculated at each depth.

\begin{figure*}
 \centering
 \includegraphics[width=14cm]{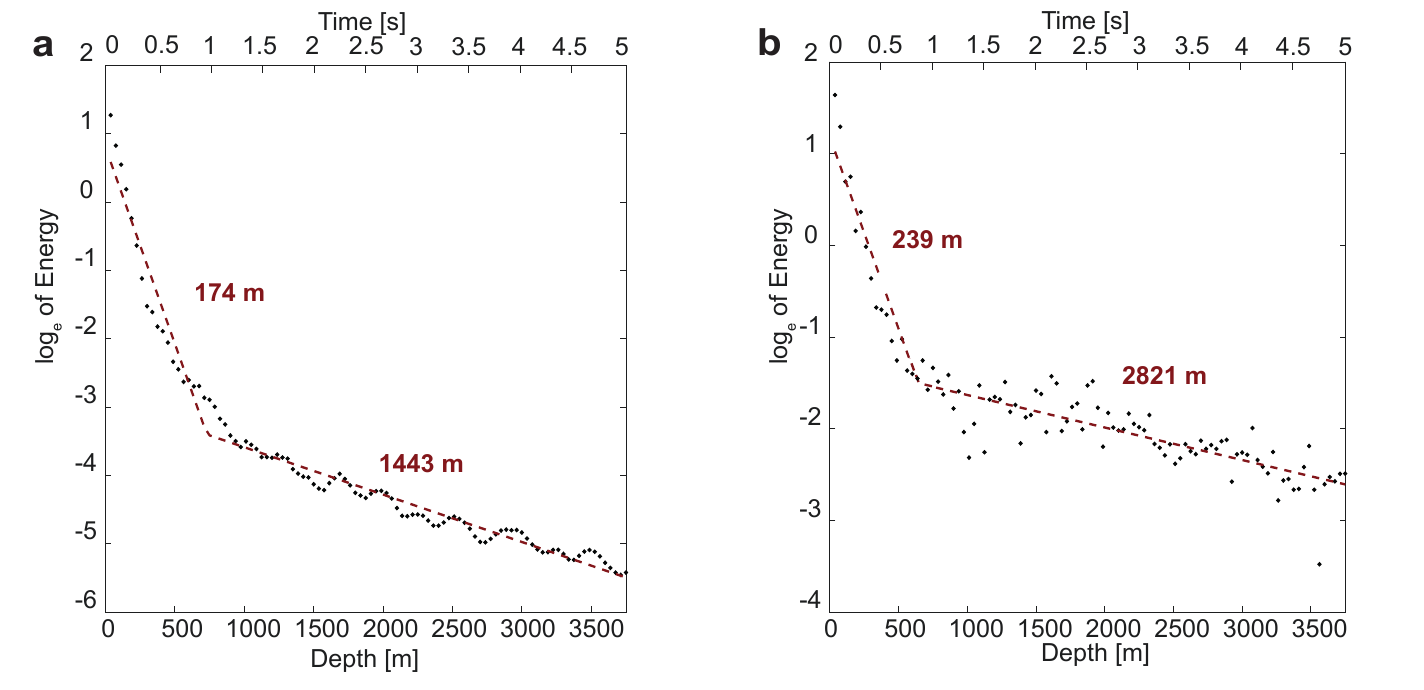}
 \caption{Mean intensity decay as a function of depth with the corresponding linear regression in logarithmic scale and the characteristic length of decay values. (a) Intensity decay of raw images. (b) Intensity decay of images corrected from the aberrating component.}
\label{figS4}
\end{figure*}

An estimation of the seismic wave attenuation at the SGB site can be directly measured through  the backscattered intensity at each depth. The energy is obtained by calculating the mean square of the intensity of the image pixels, i.e. the intensity of the diagonal of the reflection matrix. The amplitude decay of the energy is expected to follow an exponential decaying as a function of depth (Fig.~\ref{figS4}).

%\begin{equation}
%\label{decay}
%{|\mathbf{I}(\mathbf{z})| \propto| %\mathbf{I}_\textrm{0}(\mathbf{z})}|\mathbf{e}^{\frac{-2\mathbf{z}}{\ell_e_x_t}},
%\end{equation}
%where $ {I_0(z)}$ is the intensity at $\mathbf{z}=0$.

Figs.~\ref{figS4}a and b displays the natural logarithm  applied to the mean intensity as a function of depth calculated from the raw reflection matrix and the corrected reflection matrix respectively (aberration correction process disclosed in~\cite{touma2021}). In the logarithmic scale, the mean confocal energy is expressed as a decreasing linear line. The characteristic length of attenuation, that corresponds to the inverse of the slope of the log-energy decay, is also shown.

%The extinction length can be estimated by regression applied to the decay function $ \ln{I(z)}$ (dashed lines in Fig.~\ref{fig4}). The slope of the linear regression gives the value of the extinction length. 
A change of slope is observed in both plots corresponding to depth of $750$ m ($t=1$ s). While a pronounced decrease in intensity occurs until $750$ m, a more gentle slope is noticed below $750$ m.
The presence of stratigraphic boundaries, absorbing sediments and intense fracturing is a possible cause of the high attenuation observed at shallow layers. 
Another thing to notice is the increase by around twice the characteristic length values after correction. We recall that the energy decay describes the losses that the seismic waves undergo while propagating inside the medium. 
In case of strong inhomogeneities and velocity variation, the wavefield is heavily distorted and the focusing inside the medium fails. The increase in the slope provided by the aberration correction process indicates an improvement in the focusing operation and consequently an enhancement of the detection.

\section{Supplementary Data} \label{supp}

\rita{Supplementary data associated with this article can be found in the online version.}

\rita{Animation D1: \href{https://ars.els-cdn.com/content/image/1-s2.0-S0012821X21005604-mmc1.mp4
}{Animated vertical slices oriented perpendicular to the fault traces.}}

\rita{Animation D2: \href{https://ars.els-cdn.com/content/image/1-s2.0-S0012821X21005604-mmc2.mp4
}{Animated vertical slices oriented parallel to the fault traces.}}

%\bibliographystyle{elsarticle-harv} 
%\bibliographystyle{elsarticle-num} 

%\bibliography{biblio2}

\begin{thebibliography}{58}%
\makeatletter
\providecommand \@ifxundefined [1]{%
 \@ifx{#1\undefined}
}%
\providecommand \@ifnum [1]{%
 \ifnum #1\expandafter \@firstoftwo
 \else \expandafter \@secondoftwo
 \fi
}%
\providecommand \@ifx [1]{%
 \ifx #1\expandafter \@firstoftwo
 \else \expandafter \@secondoftwo
 \fi
}%
\providecommand \natexlab [1]{#1}%
\providecommand \enquote  [1]{``#1''}%
\providecommand \bibnamefont  [1]{#1}%
\providecommand \bibfnamefont [1]{#1}%
\providecommand \citenamefont [1]{#1}%
\providecommand \href@noop [0]{\@secondoftwo}%
\providecommand \href [0]{\begingroup \@sanitize@url \@href}%
\providecommand \@href[1]{\@@startlink{#1}\@@href}%
\providecommand \@@href[1]{\endgroup#1\@@endlink}%
\providecommand \@sanitize@url [0]{\catcode `\\12\catcode `\$12\catcode
  `\&12\catcode `\#12\catcode `\^12\catcode `\_12\catcode `\%12\relax}%
\providecommand \@@startlink[1]{}%
\providecommand \@@endlink[0]{}%
\providecommand \url  [0]{\begingroup\@sanitize@url \@url }%
\providecommand \@url [1]{\endgroup\@href {#1}{\urlprefix }}%
\providecommand \urlprefix  [0]{URL }%
\providecommand \Eprint [0]{\href }%
\providecommand \doibase [0]{https://doi.org/}%
\providecommand \selectlanguage [0]{\@gobble}%
\providecommand \bibinfo  [0]{\@secondoftwo}%
\providecommand \bibfield  [0]{\@secondoftwo}%
\providecommand \translation [1]{[#1]}%
\providecommand \BibitemOpen [0]{}%
\providecommand \bibitemStop [0]{}%
\providecommand \bibitemNoStop [0]{.\EOS\space}%
\providecommand \EOS [0]{\spacefactor3000\relax}%
\providecommand \BibitemShut  [1]{\csname bibitem#1\endcsname}%
\let\auto@bib@innerbib\@empty
%</preamble>
\bibitem [{\citenamefont {Touma}\ \emph {et~al.}(2021)\citenamefont {Touma},
  \citenamefont {Blondel}, \citenamefont {Derode}, \citenamefont {Campillo},\
  and\ \citenamefont {Aubry}}]{touma2021}%
  \BibitemOpen
  \bibfield  {author} {\bibinfo {author} {\bibfnamefont {R.}~\bibnamefont
  {Touma}}, \bibinfo {author} {\bibfnamefont {T.}~\bibnamefont {Blondel}},
  \bibinfo {author} {\bibfnamefont {A.}~\bibnamefont {Derode}}, \bibinfo
  {author} {\bibfnamefont {M.}~\bibnamefont {Campillo}},\ and\ \bibinfo
  {author} {\bibfnamefont {A.}~\bibnamefont {Aubry}},\ }\bibfield  {title}
  {\bibinfo {title} {A distortion matrix framework for high-resolution passive
  seismic 3-d imaging: application to the san jacinto fault zone, california},\
  }\href@noop {} {\bibfield  {journal} {\bibinfo  {journal} {Geophysical
  Journal International}\ }\textbf {\bibinfo {volume} {226}},\ \bibinfo {pages}
  {780} (\bibinfo {year} {2021})}\BibitemShut {NoStop}%
\bibitem [{\citenamefont {Chester}\ \emph {et~al.}(1993)\citenamefont
  {Chester}, \citenamefont {Evans},\ and\ \citenamefont
  {Biegel}}]{chester1993}%
  \BibitemOpen
  \bibfield  {author} {\bibinfo {author} {\bibfnamefont {F.~M.}\ \bibnamefont
  {Chester}}, \bibinfo {author} {\bibfnamefont {J.~P.}\ \bibnamefont {Evans}},\
  and\ \bibinfo {author} {\bibfnamefont {R.~L.}\ \bibnamefont {Biegel}},\
  }\bibfield  {title} {\bibinfo {title} {Internal structure and weakening
  mechanisms of the san andreas fault},\ }\href@noop {} {\bibfield  {journal}
  {\bibinfo  {journal} {J. Geophys. Res.: Solid Earth}\ }\textbf {\bibinfo
  {volume} {98}},\ \bibinfo {pages} {771} (\bibinfo {year} {1993})}\BibitemShut
  {NoStop}%
\bibitem [{\citenamefont {Wesnousky}(1988)}]{wesnousky1988}%
  \BibitemOpen
  \bibfield  {author} {\bibinfo {author} {\bibfnamefont {S.~G.}\ \bibnamefont
  {Wesnousky}},\ }\bibfield  {title} {\bibinfo {title} {Seismological and
  structural evolution of strike-slip faults},\ }\href@noop {} {\bibfield
  {journal} {\bibinfo  {journal} {Nature}\ }\textbf {\bibinfo {volume} {335}},\
  \bibinfo {pages} {340} (\bibinfo {year} {1988})}\BibitemShut {NoStop}%
\bibitem [{\citenamefont {Ben-Zion}(2008)}]{BZ2008}%
  \BibitemOpen
  \bibfield  {author} {\bibinfo {author} {\bibfnamefont {Y.}~\bibnamefont
  {Ben-Zion}},\ }\bibfield  {title} {\bibinfo {title} {Collective behavior of
  earthquakes and faults: Continuum-discrete transitions, progressive
  evolutionary changes, and different dynamic regimes},\ }\href@noop {}
  {\bibfield  {journal} {\bibinfo  {journal} {Rev. of Geophys.}\ }\textbf
  {\bibinfo {volume} {46}} (\bibinfo {year} {2008})}\BibitemShut {NoStop}%
\bibitem [{\citenamefont {Mitchell}\ and\ \citenamefont
  {Faulkner}(2009)}]{mitchell2009}%
  \BibitemOpen
  \bibfield  {author} {\bibinfo {author} {\bibfnamefont {T.}~\bibnamefont
  {Mitchell}}\ and\ \bibinfo {author} {\bibfnamefont {D.}~\bibnamefont
  {Faulkner}},\ }\bibfield  {title} {\bibinfo {title} {The nature and origin of
  off-fault damage surrounding strike-slip fault zones with a wide range of
  displacements: A field study from the atacama fault system, northern chile},\
  }\href@noop {} {\bibfield  {journal} {\bibinfo  {journal} {Jour. of Struct.
  Geo.}\ }\textbf {\bibinfo {volume} {31}},\ \bibinfo {pages} {802} (\bibinfo
  {year} {2009})}\BibitemShut {NoStop}%
\bibitem [{\citenamefont {Ben-Zion}\ and\ \citenamefont
  {Sammis}(2003)}]{BZandSammis2003}%
  \BibitemOpen
  \bibfield  {author} {\bibinfo {author} {\bibfnamefont {Y.}~\bibnamefont
  {Ben-Zion}}\ and\ \bibinfo {author} {\bibfnamefont {C.~G.}\ \bibnamefont
  {Sammis}},\ }\bibfield  {title} {\bibinfo {title} {Characterization of fault
  zones},\ }\href@noop {} {\bibfield  {journal} {\bibinfo  {journal} {Pure and
  applied geophysics}\ }\textbf {\bibinfo {volume} {160}},\ \bibinfo {pages}
  {677} (\bibinfo {year} {2003})}\BibitemShut {NoStop}%
\bibitem [{\citenamefont {Knipe}\ \emph {et~al.}(1998)\citenamefont {Knipe},
  \citenamefont {Jones},\ and\ \citenamefont {Fisher}}]{knipe1998}%
  \BibitemOpen
  \bibfield  {author} {\bibinfo {author} {\bibfnamefont {R.~J.}\ \bibnamefont
  {Knipe}}, \bibinfo {author} {\bibfnamefont {G.}~\bibnamefont {Jones}},\ and\
  \bibinfo {author} {\bibfnamefont {Q.}~\bibnamefont {Fisher}},\ }\bibfield
  {title} {\bibinfo {title} {Faulting, fault sealing and fluid flow in
  hydrocarbon reservoirs: an introduction},\ }\href@noop {} {\bibfield
  {journal} {\bibinfo  {journal} {Geological Society, London, Special
  Publications}\ }\textbf {\bibinfo {volume} {147}},\ \bibinfo {pages} {vii}
  (\bibinfo {year} {1998})}\BibitemShut {NoStop}%
\bibitem [{\citenamefont {Massonnet}\ \emph {et~al.}(1993)\citenamefont
  {Massonnet}, \citenamefont {Rossi}, \citenamefont {Carmona}, \citenamefont
  {Adragna}, \citenamefont {Peltzer}, \citenamefont {Feigl},\ and\
  \citenamefont {Rabaute}}]{massonnet1993}%
  \BibitemOpen
  \bibfield  {author} {\bibinfo {author} {\bibfnamefont {D.}~\bibnamefont
  {Massonnet}}, \bibinfo {author} {\bibfnamefont {M.}~\bibnamefont {Rossi}},
  \bibinfo {author} {\bibfnamefont {C.}~\bibnamefont {Carmona}}, \bibinfo
  {author} {\bibfnamefont {F.}~\bibnamefont {Adragna}}, \bibinfo {author}
  {\bibfnamefont {G.}~\bibnamefont {Peltzer}}, \bibinfo {author} {\bibfnamefont
  {K.}~\bibnamefont {Feigl}},\ and\ \bibinfo {author} {\bibfnamefont
  {T.}~\bibnamefont {Rabaute}},\ }\bibfield  {title} {\bibinfo {title} {The
  displacement field of the landers earthquake mapped by radar
  interferometry},\ }\href@noop {} {\bibfield  {journal} {\bibinfo  {journal}
  {Nature}\ }\textbf {\bibinfo {volume} {364}},\ \bibinfo {pages} {138}
  (\bibinfo {year} {1993})}\BibitemShut {NoStop}%
\bibitem [{\citenamefont {Binet}\ and\ \citenamefont
  {Bollinger}(2005)}]{binet2005}%
  \BibitemOpen
  \bibfield  {author} {\bibinfo {author} {\bibfnamefont {R.}~\bibnamefont
  {Binet}}\ and\ \bibinfo {author} {\bibfnamefont {L.}~\bibnamefont
  {Bollinger}},\ }\bibfield  {title} {\bibinfo {title} {Horizontal coseismic
  deformation of the 2003 bam (iran) earthquake measured from spot-5 thr
  satellite imagery},\ }\href@noop {} {\bibfield  {journal} {\bibinfo
  {journal} {Geophys. Res. Let.}\ }\textbf {\bibinfo {volume} {32}} (\bibinfo
  {year} {2005})}\BibitemShut {NoStop}%
\bibitem [{\citenamefont {Rockwell}\ and\ \citenamefont
  {Ben-Zion}(2007)}]{rockwell2007}%
  \BibitemOpen
  \bibfield  {author} {\bibinfo {author} {\bibfnamefont {T.~K.}\ \bibnamefont
  {Rockwell}}\ and\ \bibinfo {author} {\bibfnamefont {Y.}~\bibnamefont
  {Ben-Zion}},\ }\bibfield  {title} {\bibinfo {title} {High localization of
  primary slip zones in large earthquakes from paleoseismic trenches:
  Observations and implications for earthquake physics},\ }\href@noop {}
  {\bibfield  {journal} {\bibinfo  {journal} {J. Geophys. Res.: Solid Earth}\
  }\textbf {\bibinfo {volume} {112}} (\bibinfo {year} {2007})}\BibitemShut
  {NoStop}%
\bibitem [{\citenamefont {Lewis}\ \emph {et~al.}(2005)\citenamefont {Lewis},
  \citenamefont {Peng}, \citenamefont {Ben-Zion},\ and\ \citenamefont
  {Vernon}}]{lewis2005}%
  \BibitemOpen
  \bibfield  {author} {\bibinfo {author} {\bibfnamefont {M.}~\bibnamefont
  {Lewis}}, \bibinfo {author} {\bibfnamefont {Z.}~\bibnamefont {Peng}},
  \bibinfo {author} {\bibfnamefont {Y.}~\bibnamefont {Ben-Zion}},\ and\
  \bibinfo {author} {\bibfnamefont {F.}~\bibnamefont {Vernon}},\ }\bibfield
  {title} {\bibinfo {title} {Shallow seismic trapping structure in the san
  jacinto fault zone near anza, california},\ }\href@noop {} {\bibfield
  {journal} {\bibinfo  {journal} {Geophys. J. Int}\ }\textbf {\bibinfo {volume}
  {162}},\ \bibinfo {pages} {867} (\bibinfo {year} {2005})}\BibitemShut
  {NoStop}%
\bibitem [{\citenamefont {Dor}\ \emph {et~al.}(2006)\citenamefont {Dor},
  \citenamefont {Rockwell},\ and\ \citenamefont {Ben-Zion}}]{dor2006}%
  \BibitemOpen
  \bibfield  {author} {\bibinfo {author} {\bibfnamefont {O.}~\bibnamefont
  {Dor}}, \bibinfo {author} {\bibfnamefont {T.~K.}\ \bibnamefont {Rockwell}},\
  and\ \bibinfo {author} {\bibfnamefont {Y.}~\bibnamefont {Ben-Zion}},\
  }\bibfield  {title} {\bibinfo {title} {Geological observations of damage
  asymmetry in the structure of the san jacinto, san andreas and punchbowl
  faults in southern california: A possible indicator for preferred rupture
  propagation direction},\ }\href@noop {} {\bibfield  {journal} {\bibinfo
  {journal} {Pure and App. Geophy.}\ }\textbf {\bibinfo {volume} {163}},\
  \bibinfo {pages} {301} (\bibinfo {year} {2006})}\BibitemShut {NoStop}%
\bibitem [{\citenamefont {Manighetti}\ \emph {et~al.}(2005)\citenamefont
  {Manighetti}, \citenamefont {Campillo}, \citenamefont {Sammis}, \citenamefont
  {Mai},\ and\ \citenamefont {King}}]{manighetti2005}%
  \BibitemOpen
  \bibfield  {author} {\bibinfo {author} {\bibfnamefont {I.}~\bibnamefont
  {Manighetti}}, \bibinfo {author} {\bibfnamefont {M.}~\bibnamefont
  {Campillo}}, \bibinfo {author} {\bibfnamefont {C.}~\bibnamefont {Sammis}},
  \bibinfo {author} {\bibfnamefont {P.}~\bibnamefont {Mai}},\ and\ \bibinfo
  {author} {\bibfnamefont {G.}~\bibnamefont {King}},\ }\bibfield  {title}
  {\bibinfo {title} {Evidence for self-similar, triangular slip distributions
  on earthquakes: Implications for earthquake and fault mechanics},\
  }\href@noop {} {\bibfield  {journal} {\bibinfo  {journal} {J. Geophys. Res.:
  Solid Earth}\ }\textbf {\bibinfo {volume} {110}} (\bibinfo {year}
  {2005})}\BibitemShut {NoStop}%
\bibitem [{\citenamefont {Xu}\ \emph {et~al.}(2012)\citenamefont {Xu},
  \citenamefont {Ben-Zion},\ and\ \citenamefont {Ampuero}}]{xu2012}%
  \BibitemOpen
  \bibfield  {author} {\bibinfo {author} {\bibfnamefont {S.}~\bibnamefont
  {Xu}}, \bibinfo {author} {\bibfnamefont {Y.}~\bibnamefont {Ben-Zion}},\ and\
  \bibinfo {author} {\bibfnamefont {J.-P.}\ \bibnamefont {Ampuero}},\
  }\bibfield  {title} {\bibinfo {title} {Properties of inelastic yielding zones
  generated by in-plane dynamic ruptures—ii. detailed parameter-space
  study},\ }\href@noop {} {\bibfield  {journal} {\bibinfo  {journal} {Geophys.
  J. Int}\ }\textbf {\bibinfo {volume} {191}},\ \bibinfo {pages} {1343}
  (\bibinfo {year} {2012})}\BibitemShut {NoStop}%
\bibitem [{\citenamefont {Etgen}\ \emph {et~al.}(2009)\citenamefont {Etgen},
  \citenamefont {Gray},\ and\ \citenamefont {Zhang}}]{etgen2009}%
  \BibitemOpen
  \bibfield  {author} {\bibinfo {author} {\bibfnamefont {J.}~\bibnamefont
  {Etgen}}, \bibinfo {author} {\bibfnamefont {S.~H.}\ \bibnamefont {Gray}},\
  and\ \bibinfo {author} {\bibfnamefont {Y.}~\bibnamefont {Zhang}},\ }\bibfield
   {title} {\bibinfo {title} {An overview of depth imaging in exploration
  geophysics},\ }\href@noop {} {\bibfield  {journal} {\bibinfo  {journal}
  {Geophysics}\ }\textbf {\bibinfo {volume} {74}},\ \bibinfo {pages} {WCA5}
  (\bibinfo {year} {2009})}\BibitemShut {NoStop}%
\bibitem [{\citenamefont {Moser}\ and\ \citenamefont
  {Howard}(2008)}]{moser2008}%
  \BibitemOpen
  \bibfield  {author} {\bibinfo {author} {\bibfnamefont {T.}~\bibnamefont
  {Moser}}\ and\ \bibinfo {author} {\bibfnamefont {C.}~\bibnamefont {Howard}},\
  }\bibfield  {title} {\bibinfo {title} {Diffraction imaging in depth},\
  }\href@noop {} {\bibfield  {journal} {\bibinfo  {journal} {Geophysical
  Prospecting}\ }\textbf {\bibinfo {volume} {56}},\ \bibinfo {pages} {627}
  (\bibinfo {year} {2008})}\BibitemShut {NoStop}%
\bibitem [{\citenamefont {Campillo}\ and\ \citenamefont
  {Roux}(2014)}]{campilloandroux2014}%
  \BibitemOpen
  \bibfield  {author} {\bibinfo {author} {\bibfnamefont {M.}~\bibnamefont
  {Campillo}}\ and\ \bibinfo {author} {\bibfnamefont {P.}~\bibnamefont
  {Roux}},\ }\bibfield  {title} {\bibinfo {title} {Seismic imaging and
  monitoring with ambient noise correlations},\ }\href@noop {} {\bibfield
  {journal} {\bibinfo  {journal} {Treatise on Geophysics}\ }\textbf {\bibinfo
  {volume} {1}},\ \bibinfo {pages} {256} (\bibinfo {year} {2014})}\BibitemShut
  {NoStop}%
\bibitem [{\citenamefont {Wapenaar}\ \emph {et~al.}(2010)\citenamefont
  {Wapenaar}, \citenamefont {Draganov}, \citenamefont {Snieder}, \citenamefont
  {Campman},\ and\ \citenamefont {Verdel}}]{wapenaar2010}%
  \BibitemOpen
  \bibfield  {author} {\bibinfo {author} {\bibfnamefont {K.}~\bibnamefont
  {Wapenaar}}, \bibinfo {author} {\bibfnamefont {D.}~\bibnamefont {Draganov}},
  \bibinfo {author} {\bibfnamefont {R.}~\bibnamefont {Snieder}}, \bibinfo
  {author} {\bibfnamefont {X.}~\bibnamefont {Campman}},\ and\ \bibinfo {author}
  {\bibfnamefont {A.}~\bibnamefont {Verdel}},\ }\bibfield  {title} {\bibinfo
  {title} {Tutorial on seismic interferometry: Part 1—basic principles and
  applications},\ }\href@noop {} {\bibfield  {journal} {\bibinfo  {journal}
  {Geophysics}\ }\textbf {\bibinfo {volume} {75}},\ \bibinfo {pages} {75A195}
  (\bibinfo {year} {2010})}\BibitemShut {NoStop}%
\bibitem [{\citenamefont {Shapiro}\ and\ \citenamefont
  {Campillo}(2004)}]{shapiro2004}%
  \BibitemOpen
  \bibfield  {author} {\bibinfo {author} {\bibfnamefont {N.~M.}\ \bibnamefont
  {Shapiro}}\ and\ \bibinfo {author} {\bibfnamefont {M.}~\bibnamefont
  {Campillo}},\ }\bibfield  {title} {\bibinfo {title} {Emergence of broadband
  rayleigh waves from correlations of the ambient seismic noise},\ }\href@noop
  {} {\bibfield  {journal} {\bibinfo  {journal} {Geophys. Res. Let.}\ }\textbf
  {\bibinfo {volume} {31}} (\bibinfo {year} {2004})}\BibitemShut {NoStop}%
\bibitem [{\citenamefont {Poli}\ \emph
  {et~al.}(2012{\natexlab{a}})\citenamefont {Poli}, \citenamefont {Pedersen},\
  and\ \citenamefont {Campillo}}]{poli2012a}%
  \BibitemOpen
  \bibfield  {author} {\bibinfo {author} {\bibfnamefont {P.}~\bibnamefont
  {Poli}}, \bibinfo {author} {\bibfnamefont {H.}~\bibnamefont {Pedersen}},\
  and\ \bibinfo {author} {\bibfnamefont {M.}~\bibnamefont {Campillo}},\
  }\bibfield  {title} {\bibinfo {title} {Emergence of body waves from
  cross-correlation of short period seismic noise},\ }\href@noop {} {\bibfield
  {journal} {\bibinfo  {journal} {Geophys. J. Int.}\ }\textbf {\bibinfo
  {volume} {188}},\ \bibinfo {pages} {549} (\bibinfo {year}
  {2012}{\natexlab{a}})}\BibitemShut {NoStop}%
\bibitem [{\citenamefont {Draganov}\ \emph {et~al.}(2007)\citenamefont
  {Draganov}, \citenamefont {Wapenaar}, \citenamefont {Mulder}, \citenamefont
  {Singer},\ and\ \citenamefont {Verdel}}]{draganov2007}%
  \BibitemOpen
  \bibfield  {author} {\bibinfo {author} {\bibfnamefont {D.}~\bibnamefont
  {Draganov}}, \bibinfo {author} {\bibfnamefont {K.}~\bibnamefont {Wapenaar}},
  \bibinfo {author} {\bibfnamefont {W.}~\bibnamefont {Mulder}}, \bibinfo
  {author} {\bibfnamefont {J.}~\bibnamefont {Singer}},\ and\ \bibinfo {author}
  {\bibfnamefont {A.}~\bibnamefont {Verdel}},\ }\bibfield  {title} {\bibinfo
  {title} {Retrieval of reflections from seismic background-noise
  measurements},\ }\href@noop {} {\bibfield  {journal} {\bibinfo  {journal}
  {Geophys. Res. Lett.}\ }\textbf {\bibinfo {volume} {34}} (\bibinfo {year}
  {2007})}\BibitemShut {NoStop}%
\bibitem [{\citenamefont {Poli}\ \emph
  {et~al.}(2012{\natexlab{b}})\citenamefont {Poli}, \citenamefont {Campillo},
  \citenamefont {Pedersen}, \citenamefont {Group} \emph {et~al.}}]{poli2012b}%
  \BibitemOpen
  \bibfield  {author} {\bibinfo {author} {\bibfnamefont {P.}~\bibnamefont
  {Poli}}, \bibinfo {author} {\bibfnamefont {M.}~\bibnamefont {Campillo}},
  \bibinfo {author} {\bibfnamefont {H.}~\bibnamefont {Pedersen}}, \bibinfo
  {author} {\bibfnamefont {L.~W.}\ \bibnamefont {Group}}, \emph {et~al.},\
  }\bibfield  {title} {\bibinfo {title} {Body-wave imaging of earth’s mantle
  discontinuities from ambient seismic noise},\ }\href@noop {} {\bibfield
  {journal} {\bibinfo  {journal} {Science}\ }\textbf {\bibinfo {volume}
  {338}},\ \bibinfo {pages} {1063} (\bibinfo {year}
  {2012}{\natexlab{b}})}\BibitemShut {NoStop}%
\bibitem [{\citenamefont {Aubry}\ and\ \citenamefont
  {Derode}(2009)}]{aubry2009}%
  \BibitemOpen
  \bibfield  {author} {\bibinfo {author} {\bibfnamefont {A.}~\bibnamefont
  {Aubry}}\ and\ \bibinfo {author} {\bibfnamefont {A.}~\bibnamefont {Derode}},\
  }\bibfield  {title} {\bibinfo {title} {Detection and imaging in a random
  medium: A matrix method to overcome multiple scattering and aberration},\
  }\href@noop {} {\bibfield  {journal} {\bibinfo  {journal} {Journal of Applied
  Physics}\ }\textbf {\bibinfo {volume} {106}},\ \bibinfo {pages} {044903}
  (\bibinfo {year} {2009})}\BibitemShut {NoStop}%
\bibitem [{\citenamefont {Badon}\ \emph {et~al.}(2016)\citenamefont {Badon},
  \citenamefont {Li}, \citenamefont {Lerosey}, \citenamefont {Boccara},
  \citenamefont {Fink},\ and\ \citenamefont {Aubry}}]{badon2016}%
  \BibitemOpen
  \bibfield  {author} {\bibinfo {author} {\bibfnamefont {A.}~\bibnamefont
  {Badon}}, \bibinfo {author} {\bibfnamefont {D.}~\bibnamefont {Li}}, \bibinfo
  {author} {\bibfnamefont {G.}~\bibnamefont {Lerosey}}, \bibinfo {author}
  {\bibfnamefont {A.~C.}\ \bibnamefont {Boccara}}, \bibinfo {author}
  {\bibfnamefont {M.}~\bibnamefont {Fink}},\ and\ \bibinfo {author}
  {\bibfnamefont {A.}~\bibnamefont {Aubry}},\ }\bibfield  {title} {\bibinfo
  {title} {Smart optical coherence tomography for ultra-deep imaging through
  highly scattering media},\ }\href@noop {} {\bibfield  {journal} {\bibinfo
  {journal} {Sci. Adv.}\ }\textbf {\bibinfo {volume} {2}},\ \bibinfo {pages}
  {e1600370} (\bibinfo {year} {2016})}\BibitemShut {NoStop}%
\bibitem [{\citenamefont {Blondel}\ \emph {et~al.}(2018)\citenamefont
  {Blondel}, \citenamefont {Chaput}, \citenamefont {Derode}, \citenamefont
  {Campillo},\ and\ \citenamefont {Aubry}}]{blondel2018}%
  \BibitemOpen
  \bibfield  {author} {\bibinfo {author} {\bibfnamefont {T.}~\bibnamefont
  {Blondel}}, \bibinfo {author} {\bibfnamefont {J.}~\bibnamefont {Chaput}},
  \bibinfo {author} {\bibfnamefont {A.}~\bibnamefont {Derode}}, \bibinfo
  {author} {\bibfnamefont {M.}~\bibnamefont {Campillo}},\ and\ \bibinfo
  {author} {\bibfnamefont {A.}~\bibnamefont {Aubry}},\ }\bibfield  {title}
  {\bibinfo {title} {Matrix approach of seismic imaging: application to the
  {E}rebus volcano, {A}ntarctica},\ }\href@noop {} {\bibfield  {journal}
  {\bibinfo  {journal} {J. Geophys. Res.: Solid Earth}\ }\textbf {\bibinfo
  {volume} {123}},\ \bibinfo {pages} {10,936} (\bibinfo {year}
  {2018})}\BibitemShut {NoStop}%
\bibitem [{\citenamefont {Lambert}\ \emph
  {et~al.}(2020{\natexlab{a}})\citenamefont {Lambert}, \citenamefont {Cobus},
  \citenamefont {Couade}, \citenamefont {Fink},\ and\ \citenamefont
  {Aubry}}]{Lambert2020}%
  \BibitemOpen
  \bibfield  {author} {\bibinfo {author} {\bibfnamefont {W.}~\bibnamefont
  {Lambert}}, \bibinfo {author} {\bibfnamefont {L.~A.}\ \bibnamefont {Cobus}},
  \bibinfo {author} {\bibfnamefont {M.}~\bibnamefont {Couade}}, \bibinfo
  {author} {\bibfnamefont {M.}~\bibnamefont {Fink}},\ and\ \bibinfo {author}
  {\bibfnamefont {A.}~\bibnamefont {Aubry}},\ }\bibfield  {title} {\bibinfo
  {title} {Reflection matrix approach for quantitative imaging of scattering
  media},\ }\href@noop {} {\bibfield  {journal} {\bibinfo  {journal} {Physical
  Review X}\ }\textbf {\bibinfo {volume} {10}},\ \bibinfo {pages} {021048}
  (\bibinfo {year} {2020}{\natexlab{a}})}\BibitemShut {NoStop}%
\bibitem [{\citenamefont {Badon}\ \emph {et~al.}(2020)\citenamefont {Badon},
  \citenamefont {Barolle}, \citenamefont {Irsch}, \citenamefont {Boccara},
  \citenamefont {Fink},\ and\ \citenamefont {Aubry}}]{Badon2019}%
  \BibitemOpen
  \bibfield  {author} {\bibinfo {author} {\bibfnamefont {A.}~\bibnamefont
  {Badon}}, \bibinfo {author} {\bibfnamefont {V.}~\bibnamefont {Barolle}},
  \bibinfo {author} {\bibfnamefont {K.}~\bibnamefont {Irsch}}, \bibinfo
  {author} {\bibfnamefont {A.~C.}\ \bibnamefont {Boccara}}, \bibinfo {author}
  {\bibfnamefont {M.}~\bibnamefont {Fink}},\ and\ \bibinfo {author}
  {\bibfnamefont {A.}~\bibnamefont {Aubry}},\ }\bibfield  {title} {\bibinfo
  {title} {{Distortion matrix concept for deep imaging in optical coherence
  microscopy}},\ }\href@noop {} {\bibfield  {journal} {\bibinfo  {journal}
  {Sci. Adv.}\ }\textbf {\bibinfo {volume} {6}},\ \bibinfo {pages} {eaay7170}
  (\bibinfo {year} {2020})}\BibitemShut {NoStop}%
\bibitem [{\citenamefont {Lambert}\ \emph
  {et~al.}(2022{\natexlab{a}})\citenamefont {Lambert}, \citenamefont {Robin},
  \citenamefont {Cobus}, \citenamefont {Fink},\ and\ \citenamefont
  {Aubry}}]{lambert2021refle}%
  \BibitemOpen
  \bibfield  {author} {\bibinfo {author} {\bibfnamefont {W.}~\bibnamefont
  {Lambert}}, \bibinfo {author} {\bibfnamefont {J.}~\bibnamefont {Robin}},
  \bibinfo {author} {\bibfnamefont {L.~A.}\ \bibnamefont {Cobus}}, \bibinfo
  {author} {\bibfnamefont {M.}~\bibnamefont {Fink}},\ and\ \bibinfo {author}
  {\bibfnamefont {A.}~\bibnamefont {Aubry}},\ }\bibfield  {title} {\bibinfo
  {title} {Ultrasound matrix imaging. {I.} {T}he focused reflection matrix, the
  {F}-factor and the role of multiple scattering},\ }\bibfield  {journal}
  {\bibinfo  {journal} {IEEE Trans. Med. Imag. (to be published)}\ }\href
  {https://doi.org/10.1109/TMI.2022.3199498} {10.1109/TMI.2022.3199498}
  (\bibinfo {year} {2022}{\natexlab{a}})\BibitemShut {NoStop}%
\bibitem [{\citenamefont {Berkhout}\ and\ \citenamefont
  {Wapenaar}(1993)}]{Berkhout1993}%
  \BibitemOpen
  \bibfield  {author} {\bibinfo {author} {\bibfnamefont {A.~J.}\ \bibnamefont
  {Berkhout}}\ and\ \bibinfo {author} {\bibfnamefont {C.~P.~A.}\ \bibnamefont
  {Wapenaar}},\ }\bibfield  {title} {\bibinfo {title} {A unified approach to
  acoustical reflection imaging. {II: T}he inverse problem},\ }\href@noop {}
  {\bibfield  {journal} {\bibinfo  {journal} {J Acoust. Soc. Am.}\ }\textbf
  {\bibinfo {volume} {93}},\ \bibinfo {pages} {2017} (\bibinfo {year}
  {1993})}\BibitemShut {NoStop}%
\bibitem [{\citenamefont {Lambert}\ \emph
  {et~al.}(2020{\natexlab{b}})\citenamefont {Lambert}, \citenamefont {Cobus},
  \citenamefont {Frappart}, \citenamefont {Fink},\ and\ \citenamefont
  {Aubry}}]{william}%
  \BibitemOpen
  \bibfield  {author} {\bibinfo {author} {\bibfnamefont {W.}~\bibnamefont
  {Lambert}}, \bibinfo {author} {\bibfnamefont {L.~A.}\ \bibnamefont {Cobus}},
  \bibinfo {author} {\bibfnamefont {T.}~\bibnamefont {Frappart}}, \bibinfo
  {author} {\bibfnamefont {M.}~\bibnamefont {Fink}},\ and\ \bibinfo {author}
  {\bibfnamefont {A.}~\bibnamefont {Aubry}},\ }\bibfield  {title} {\bibinfo
  {title} {Distortion matrix approach for ultrasound imaging of random
  scattering media},\ }\href@noop {} {\bibfield  {journal} {\bibinfo  {journal}
  {Proc. Nat. Sci. Acad.}\ }\textbf {\bibinfo {volume} {117}},\ \bibinfo
  {pages} {14645} (\bibinfo {year} {2020}{\natexlab{b}})}\BibitemShut {NoStop}%
\bibitem [{\citenamefont {Lambert}\ \emph
  {et~al.}(2022{\natexlab{b}})\citenamefont {Lambert}, \citenamefont {Cobus},
  \citenamefont {Robin}, \citenamefont {Fink},\ and\ \citenamefont
  {Aubry}}]{lambert2021distor}%
  \BibitemOpen
  \bibfield  {author} {\bibinfo {author} {\bibfnamefont {W.}~\bibnamefont
  {Lambert}}, \bibinfo {author} {\bibfnamefont {L.~A.}\ \bibnamefont {Cobus}},
  \bibinfo {author} {\bibfnamefont {J.}~\bibnamefont {Robin}}, \bibinfo
  {author} {\bibfnamefont {M.}~\bibnamefont {Fink}},\ and\ \bibinfo {author}
  {\bibfnamefont {A.}~\bibnamefont {Aubry}},\ }\bibfield  {title} {\bibinfo
  {title} {Ultrasound matrix imaging. {II. T}he distortion matrix for a local
  correction of aberrations},\ }\bibfield  {journal} {\bibinfo  {journal} {IEEE
  Trans. Med. Imag. (to be published)}\ }\href
  {https://doi.org/10.1109/TMI.2022.3199483} {10.1109/TMI.2022.3199483}
  (\bibinfo {year} {2022}{\natexlab{b}})\BibitemShut {NoStop}%
\bibitem [{\citenamefont {Ben-Zion}\ \emph {et~al.}(2015)\citenamefont
  {Ben-Zion}, \citenamefont {Vernon}, \citenamefont {Ozakin}, \citenamefont
  {Zigone}, \citenamefont {Ross}, \citenamefont {Meng}, \citenamefont {White},
  \citenamefont {Reyes}, \citenamefont {Hollis},\ and\ \citenamefont
  {Barklage}}]{Ben_Zion_2015}%
  \BibitemOpen
  \bibfield  {author} {\bibinfo {author} {\bibfnamefont {Y.}~\bibnamefont
  {Ben-Zion}}, \bibinfo {author} {\bibfnamefont {F.~L.}\ \bibnamefont
  {Vernon}}, \bibinfo {author} {\bibfnamefont {Y.}~\bibnamefont {Ozakin}},
  \bibinfo {author} {\bibfnamefont {D.}~\bibnamefont {Zigone}}, \bibinfo
  {author} {\bibfnamefont {Z.~E.}\ \bibnamefont {Ross}}, \bibinfo {author}
  {\bibfnamefont {H.}~\bibnamefont {Meng}}, \bibinfo {author} {\bibfnamefont
  {M.}~\bibnamefont {White}}, \bibinfo {author} {\bibfnamefont
  {J.}~\bibnamefont {Reyes}}, \bibinfo {author} {\bibfnamefont
  {D.}~\bibnamefont {Hollis}},\ and\ \bibinfo {author} {\bibfnamefont
  {M.}~\bibnamefont {Barklage}},\ }\bibfield  {title} {\bibinfo {title} {Basic
  data features and results from a spatially dense seismic array on the {S}an
  {J}acinto fault zone},\ }\href {https://doi.org/10.1093/gji/ggv142}
  {\bibfield  {journal} {\bibinfo  {journal} {Geophys. J. Int.}\ }\textbf
  {\bibinfo {volume} {202}},\ \bibinfo {pages} {370} (\bibinfo {year}
  {2015})}\BibitemShut {NoStop}%
\bibitem [{\citenamefont {Roux}\ \emph {et~al.}(2016)\citenamefont {Roux},
  \citenamefont {Moreau}, \citenamefont {Lecointre}, \citenamefont {Hillers},
  \citenamefont {Campillo}, \citenamefont {Ben-Zion}, \citenamefont {Zigone},\
  and\ \citenamefont {Vernon}}]{Roux_2016}%
  \BibitemOpen
  \bibfield  {author} {\bibinfo {author} {\bibfnamefont {P.}~\bibnamefont
  {Roux}}, \bibinfo {author} {\bibfnamefont {L.}~\bibnamefont {Moreau}},
  \bibinfo {author} {\bibfnamefont {A.}~\bibnamefont {Lecointre}}, \bibinfo
  {author} {\bibfnamefont {G.}~\bibnamefont {Hillers}}, \bibinfo {author}
  {\bibfnamefont {M.}~\bibnamefont {Campillo}}, \bibinfo {author}
  {\bibfnamefont {Y.}~\bibnamefont {Ben-Zion}}, \bibinfo {author}
  {\bibfnamefont {D.}~\bibnamefont {Zigone}},\ and\ \bibinfo {author}
  {\bibfnamefont {F.}~\bibnamefont {Vernon}},\ }\bibfield  {title} {\bibinfo
  {title} {{A methodological approach towards high-resolution surface wave
  imaging of the San Jacinto Fault Zone using ambient-noise recordings at a
  spatially dense array}},\ }\href {https://doi.org/10.1093/gji/ggw193}
  {\bibfield  {journal} {\bibinfo  {journal} {Geophys. J. Int.}\ }\textbf
  {\bibinfo {volume} {206}},\ \bibinfo {pages} {980} (\bibinfo {year}
  {2016})}\BibitemShut {NoStop}%
\bibitem [{\citenamefont {Wade}(2018)}]{wade2018}%
  \BibitemOpen
  \bibfield  {author} {\bibinfo {author} {\bibfnamefont {A.~M.}\ \bibnamefont
  {Wade}},\ }\bibfield  {title} {\bibinfo {title} {Geologic and structural
  characterization of shallow seismic properties along the san jacinto fault at
  sage brush flat, southern california},\ }\href@noop {} {\bibfield  {journal}
  {\bibinfo  {journal} {Master Thesis}\ ,\ \bibinfo {pages} {8}} (\bibinfo
  {year} {2018})}\BibitemShut {NoStop}%
\bibitem [{\citenamefont {Share}\ \emph {et~al.}(2020)\citenamefont {Share},
  \citenamefont {T{\'a}bo{\v{r}}{\'\i}k}, \citenamefont
  {{\v{S}}t{\v{e}}pan{\v{c}}{\'\i}kov{\'a}}, \citenamefont {Stemberk},
  \citenamefont {Rockwell}, \citenamefont {Wade}, \citenamefont {Arrowsmith},
  \citenamefont {Donnellan}, \citenamefont {Vernon},\ and\ \citenamefont
  {Ben-Zion}}]{share2020}%
  \BibitemOpen
  \bibfield  {author} {\bibinfo {author} {\bibfnamefont {P.-E.}\ \bibnamefont
  {Share}}, \bibinfo {author} {\bibfnamefont {P.}~\bibnamefont
  {T{\'a}bo{\v{r}}{\'\i}k}}, \bibinfo {author} {\bibfnamefont {P.}~\bibnamefont
  {{\v{S}}t{\v{e}}pan{\v{c}}{\'\i}kov{\'a}}}, \bibinfo {author} {\bibfnamefont
  {J.}~\bibnamefont {Stemberk}}, \bibinfo {author} {\bibfnamefont {T.~K.}\
  \bibnamefont {Rockwell}}, \bibinfo {author} {\bibfnamefont {A.}~\bibnamefont
  {Wade}}, \bibinfo {author} {\bibfnamefont {J.~R.}\ \bibnamefont
  {Arrowsmith}}, \bibinfo {author} {\bibfnamefont {A.}~\bibnamefont
  {Donnellan}}, \bibinfo {author} {\bibfnamefont {F.~L.}\ \bibnamefont
  {Vernon}},\ and\ \bibinfo {author} {\bibfnamefont {Y.}~\bibnamefont
  {Ben-Zion}},\ }\bibfield  {title} {\bibinfo {title} {Characterizing the
  uppermost 100 m structure of the san jacinto fault zone southeast of anza,
  california, through joint analysis of geological, topographic, seismic and
  resistivity data},\ }\href@noop {} {\bibfield  {journal} {\bibinfo  {journal}
  {Geophys. J. Int}\ }\textbf {\bibinfo {volume} {222}},\ \bibinfo {pages}
  {781} (\bibinfo {year} {2020})}\BibitemShut {NoStop}%
\bibitem [{\citenamefont {Hauksson}\ \emph {et~al.}(2012)\citenamefont
  {Hauksson}, \citenamefont {Yang},\ and\ \citenamefont
  {Shearer}}]{hauksson2012}%
  \BibitemOpen
  \bibfield  {author} {\bibinfo {author} {\bibfnamefont {E.}~\bibnamefont
  {Hauksson}}, \bibinfo {author} {\bibfnamefont {W.}~\bibnamefont {Yang}},\
  and\ \bibinfo {author} {\bibfnamefont {P.~M.}\ \bibnamefont {Shearer}},\
  }\bibfield  {title} {\bibinfo {title} {Waveform relocated earthquake catalog
  for southern california (1981 to june 2011)},\ }\href@noop {} {\bibfield
  {journal} {\bibinfo  {journal} {Bulletin of the Seis. Soc. of Amer.}\
  }\textbf {\bibinfo {volume} {102}},\ \bibinfo {pages} {2239} (\bibinfo {year}
  {2012})}\BibitemShut {NoStop}%
\bibitem [{\citenamefont {Matti}\ and\ \citenamefont
  {Morton}(1993)}]{matti1993}%
  \BibitemOpen
  \bibfield  {author} {\bibinfo {author} {\bibfnamefont {J.~C.}\ \bibnamefont
  {Matti}}\ and\ \bibinfo {author} {\bibfnamefont {D.}~\bibnamefont {Morton}},\
  }\bibfield  {title} {\bibinfo {title} {Extension and contraction within an
  evolving divergent strike-slip fault complex: The san andreas and san jacinto
  fault zones at their convergence in southern california},\ }\href@noop {}
  {\bibfield  {journal} {\bibinfo  {journal} {The San Andreas fault system:
  Displacement, palinspastic reconstruction, and geologic evolution}\ }\textbf
  {\bibinfo {volume} {178}},\ \bibinfo {pages} {217} (\bibinfo {year}
  {1993})}\BibitemShut {NoStop}%
\bibitem [{\citenamefont {Dorsey}\ and\ \citenamefont
  {Roering}(2006)}]{dorsey2006}%
  \BibitemOpen
  \bibfield  {author} {\bibinfo {author} {\bibfnamefont {R.~J.}\ \bibnamefont
  {Dorsey}}\ and\ \bibinfo {author} {\bibfnamefont {J.~J.}\ \bibnamefont
  {Roering}},\ }\bibfield  {title} {\bibinfo {title} {Quaternary landscape
  evolution in the san jacinto fault zone, peninsular ranges of southern
  california: Transient response to strike-slip fault initiation},\ }\href@noop
  {} {\bibfield  {journal} {\bibinfo  {journal} {Geomorphology}\ }\textbf
  {\bibinfo {volume} {73}},\ \bibinfo {pages} {16} (\bibinfo {year}
  {2006})}\BibitemShut {NoStop}%
\bibitem [{\citenamefont {Rockwell}\ \emph {et~al.}(2003)\citenamefont
  {Rockwell}, \citenamefont {Young}, \citenamefont {Seitz}, \citenamefont
  {Meltzner}, \citenamefont {Verdugo}, \citenamefont {Khatib}, \citenamefont
  {Ragona}, \citenamefont {Altangerel},\ and\ \citenamefont
  {West}}]{rockwell2003}%
  \BibitemOpen
  \bibfield  {author} {\bibinfo {author} {\bibfnamefont {T.~K.}\ \bibnamefont
  {Rockwell}}, \bibinfo {author} {\bibfnamefont {J.}~\bibnamefont {Young}},
  \bibinfo {author} {\bibfnamefont {G.}~\bibnamefont {Seitz}}, \bibinfo
  {author} {\bibfnamefont {A.}~\bibnamefont {Meltzner}}, \bibinfo {author}
  {\bibfnamefont {D.}~\bibnamefont {Verdugo}}, \bibinfo {author} {\bibfnamefont
  {F.}~\bibnamefont {Khatib}}, \bibinfo {author} {\bibfnamefont
  {D.}~\bibnamefont {Ragona}}, \bibinfo {author} {\bibfnamefont
  {O.}~\bibnamefont {Altangerel}},\ and\ \bibinfo {author} {\bibfnamefont
  {J.}~\bibnamefont {West}},\ }\bibfield  {title} {\bibinfo {title} {3,000
  years of groundrupturing earthquakes in the anza seismic gap, san jacinto
  fault, southern california: Time to shake it up},\ }\href@noop {} {\bibfield
  {journal} {\bibinfo  {journal} {Seismol. Res. Lett}\ }\textbf {\bibinfo
  {volume} {74}},\ \bibinfo {pages} {236} (\bibinfo {year} {2003})}\BibitemShut
  {NoStop}%
\bibitem [{\citenamefont {Fialko}(2006)}]{fialko2006}%
  \BibitemOpen
  \bibfield  {author} {\bibinfo {author} {\bibfnamefont {Y.}~\bibnamefont
  {Fialko}},\ }\bibfield  {title} {\bibinfo {title} {Interseismic strain
  accumulation and the earthquake potential on the southern san andreas fault
  system},\ }\href@noop {} {\bibfield  {journal} {\bibinfo  {journal} {Nature}\
  }\textbf {\bibinfo {volume} {441}},\ \bibinfo {pages} {968} (\bibinfo {year}
  {2006})}\BibitemShut {NoStop}%
\bibitem [{\citenamefont {Sanders}\ and\ \citenamefont
  {Kanamori}(1984)}]{sanders1984}%
  \BibitemOpen
  \bibfield  {author} {\bibinfo {author} {\bibfnamefont {C.~O.}\ \bibnamefont
  {Sanders}}\ and\ \bibinfo {author} {\bibfnamefont {H.}~\bibnamefont
  {Kanamori}},\ }\bibfield  {title} {\bibinfo {title} {A seismotectonic
  analysis of the anza seismic gap, san jacinto fault zone, southern
  california},\ }\href@noop {} {\bibfield  {journal} {\bibinfo  {journal} {J.
  Geophys. Res.: Solid Earth}\ }\textbf {\bibinfo {volume} {89}},\ \bibinfo
  {pages} {5873} (\bibinfo {year} {1984})}\BibitemShut {NoStop}%
\bibitem [{\citenamefont {Ross}\ \emph {et~al.}(2017)\citenamefont {Ross},
  \citenamefont {Hauksson},\ and\ \citenamefont {Ben-Zion}}]{ross2017}%
  \BibitemOpen
  \bibfield  {author} {\bibinfo {author} {\bibfnamefont {Z.~E.}\ \bibnamefont
  {Ross}}, \bibinfo {author} {\bibfnamefont {E.}~\bibnamefont {Hauksson}},\
  and\ \bibinfo {author} {\bibfnamefont {Y.}~\bibnamefont {Ben-Zion}},\
  }\bibfield  {title} {\bibinfo {title} {Abundant off-fault seismicity and
  orthogonal structures in the san jacinto fault zone},\ }\href@noop {}
  {\bibfield  {journal} {\bibinfo  {journal} {Sci. Adv.}\ }\textbf {\bibinfo
  {volume} {3}},\ \bibinfo {pages} {e1601946} (\bibinfo {year}
  {2017})}\BibitemShut {NoStop}%
\bibitem [{\citenamefont {Khaidukov}\ \emph {et~al.}(2004)\citenamefont
  {Khaidukov}, \citenamefont {Landa},\ and\ \citenamefont
  {Moser}}]{khaidukov2004}%
  \BibitemOpen
  \bibfield  {author} {\bibinfo {author} {\bibfnamefont {V.}~\bibnamefont
  {Khaidukov}}, \bibinfo {author} {\bibfnamefont {E.}~\bibnamefont {Landa}},\
  and\ \bibinfo {author} {\bibfnamefont {T.~J.}\ \bibnamefont {Moser}},\
  }\bibfield  {title} {\bibinfo {title} {Diffraction imaging by
  focusing-defocusing: An outlook on seismic superresolution},\ }\href@noop {}
  {\bibfield  {journal} {\bibinfo  {journal} {Geophysics}\ }\textbf {\bibinfo
  {volume} {69}},\ \bibinfo {pages} {1478} (\bibinfo {year}
  {2004})}\BibitemShut {NoStop}%
\bibitem [{\citenamefont {Schwarz}(2019)}]{schwarz2019}%
  \BibitemOpen
  \bibfield  {author} {\bibinfo {author} {\bibfnamefont {B.}~\bibnamefont
  {Schwarz}},\ }\bibfield  {title} {\bibinfo {title} {An introduction to
  seismic diffraction},\ }in\ \href@noop {} {\emph {\bibinfo {booktitle}
  {Advances in Geophysics}}},\ Vol.~\bibinfo {volume} {60}\ (\bibinfo
  {publisher} {Elsevier},\ \bibinfo {year} {2019})\ pp.\ \bibinfo {pages}
  {1--64}\BibitemShut {NoStop}%
\bibitem [{\citenamefont {Kozlov}\ \emph {et~al.}(2004)\citenamefont {Kozlov},
  \citenamefont {Barasky}, \citenamefont {Korolev}, \citenamefont {Antonenko},\
  and\ \citenamefont {Koshchuk}}]{kozlov2004}%
  \BibitemOpen
  \bibfield  {author} {\bibinfo {author} {\bibfnamefont {E.}~\bibnamefont
  {Kozlov}}, \bibinfo {author} {\bibfnamefont {N.}~\bibnamefont {Barasky}},
  \bibinfo {author} {\bibfnamefont {E.}~\bibnamefont {Korolev}}, \bibinfo
  {author} {\bibfnamefont {A.}~\bibnamefont {Antonenko}},\ and\ \bibinfo
  {author} {\bibfnamefont {E.}~\bibnamefont {Koshchuk}},\ }\bibfield  {title}
  {\bibinfo {title} {Imaging scattering objects masked by specular
  reflections},\ }in\ \href@noop {} {\emph {\bibinfo {booktitle} {SEG Technical
  Program Expanded Abstracts 2004}}}\ (\bibinfo  {publisher} {Society of
  Exploration Geophysicists},\ \bibinfo {year} {2004})\ pp.\ \bibinfo {pages}
  {1131--1134}\BibitemShut {NoStop}%
\bibitem [{\citenamefont {Schwarz}\ and\ \citenamefont
  {Krawczyk}(2020)}]{schwarz2020}%
  \BibitemOpen
  \bibfield  {author} {\bibinfo {author} {\bibfnamefont {B.}~\bibnamefont
  {Schwarz}}\ and\ \bibinfo {author} {\bibfnamefont {C.~M.}\ \bibnamefont
  {Krawczyk}},\ }\bibfield  {title} {\bibinfo {title} {Coherent diffraction
  imaging for enhanced fault and fracture network characterization},\
  }\href@noop {} {\bibfield  {journal} {\bibinfo  {journal} {Solid Earth}\
  }\textbf {\bibinfo {volume} {11}},\ \bibinfo {pages} {1891} (\bibinfo {year}
  {2020})}\BibitemShut {NoStop}%
\bibitem [{\citenamefont {Kanasewich}\ and\ \citenamefont
  {Phadke}(1988)}]{kanasewich1988}%
  \BibitemOpen
  \bibfield  {author} {\bibinfo {author} {\bibfnamefont {E.~R.}\ \bibnamefont
  {Kanasewich}}\ and\ \bibinfo {author} {\bibfnamefont {S.~M.}\ \bibnamefont
  {Phadke}},\ }\bibfield  {title} {\bibinfo {title} {Imaging discontinuities on
  seismic sections},\ }\href@noop {} {\bibfield  {journal} {\bibinfo  {journal}
  {Geophysics}\ }\textbf {\bibinfo {volume} {53}},\ \bibinfo {pages} {334}
  (\bibinfo {year} {1988})}\BibitemShut {NoStop}%
\bibitem [{\citenamefont {Bakhtiari~Rad}\ \emph {et~al.}(2018)\citenamefont
  {Bakhtiari~Rad}, \citenamefont {Schwarz}, \citenamefont {Gajewski},\ and\
  \citenamefont {Vanelle}}]{bakhtiari2018}%
  \BibitemOpen
  \bibfield  {author} {\bibinfo {author} {\bibfnamefont {P.}~\bibnamefont
  {Bakhtiari~Rad}}, \bibinfo {author} {\bibfnamefont {B.}~\bibnamefont
  {Schwarz}}, \bibinfo {author} {\bibfnamefont {D.}~\bibnamefont {Gajewski}},\
  and\ \bibinfo {author} {\bibfnamefont {C.}~\bibnamefont {Vanelle}},\
  }\bibfield  {title} {\bibinfo {title} {Common-reflection-surface-based
  prestack diffraction separation and imaging},\ }\href@noop {} {\bibfield
  {journal} {\bibinfo  {journal} {Geophysics}\ }\textbf {\bibinfo {volume}
  {83}},\ \bibinfo {pages} {S47} (\bibinfo {year} {2018})}\BibitemShut
  {NoStop}%
\bibitem [{\citenamefont {Hillers}\ \emph {et~al.}(2016)\citenamefont
  {Hillers}, \citenamefont {Roux}, \citenamefont {Campillo},\ and\
  \citenamefont {Ben-Zion}}]{hillers2016}%
  \BibitemOpen
  \bibfield  {author} {\bibinfo {author} {\bibfnamefont {G.}~\bibnamefont
  {Hillers}}, \bibinfo {author} {\bibfnamefont {P.}~\bibnamefont {Roux}},
  \bibinfo {author} {\bibfnamefont {M.}~\bibnamefont {Campillo}},\ and\
  \bibinfo {author} {\bibfnamefont {Y.}~\bibnamefont {Ben-Zion}},\ }\bibfield
  {title} {\bibinfo {title} {{Focal spot imaging based on zero lag
  cross-correlation amplitude fields: Application to dense array data at the
  San Jacinto fault zone}},\ }\href@noop {} {\bibfield  {journal} {\bibinfo
  {journal} {J. Geophys. Res.: Solid Earth}\ }\textbf {\bibinfo {volume}
  {121}},\ \bibinfo {pages} {8048} (\bibinfo {year} {2016})}\BibitemShut
  {NoStop}%
\bibitem [{\citenamefont {Mordret}\ \emph {et~al.}(2019)\citenamefont
  {Mordret}, \citenamefont {Roux}, \citenamefont {Bou{\'e}},\ and\
  \citenamefont {Ben-Zion}}]{mordret2019}%
  \BibitemOpen
  \bibfield  {author} {\bibinfo {author} {\bibfnamefont {A.}~\bibnamefont
  {Mordret}}, \bibinfo {author} {\bibfnamefont {P.}~\bibnamefont {Roux}},
  \bibinfo {author} {\bibfnamefont {P.}~\bibnamefont {Bou{\'e}}},\ and\
  \bibinfo {author} {\bibfnamefont {Y.}~\bibnamefont {Ben-Zion}},\ }\bibfield
  {title} {\bibinfo {title} {{Shallow three-dimensional structure of the San
  Jacinto fault zone revealed from ambient noise imaging with a dense seismic
  array}},\ }\href@noop {} {\bibfield  {journal} {\bibinfo  {journal} {Geophys.
  J. Int.}\ }\textbf {\bibinfo {volume} {216}},\ \bibinfo {pages} {896}
  (\bibinfo {year} {2019})}\BibitemShut {NoStop}%
\bibitem [{\citenamefont {Qin}\ \emph {et~al.}(2018)\citenamefont {Qin},
  \citenamefont {Ben-Zion}, \citenamefont {Qiu}, \citenamefont {Share},
  \citenamefont {Ross},\ and\ \citenamefont {Vernon}}]{qin2018}%
  \BibitemOpen
  \bibfield  {author} {\bibinfo {author} {\bibfnamefont {L.}~\bibnamefont
  {Qin}}, \bibinfo {author} {\bibfnamefont {Y.}~\bibnamefont {Ben-Zion}},
  \bibinfo {author} {\bibfnamefont {H.}~\bibnamefont {Qiu}}, \bibinfo {author}
  {\bibfnamefont {P.}~\bibnamefont {Share}}, \bibinfo {author} {\bibfnamefont
  {Z.}~\bibnamefont {Ross}},\ and\ \bibinfo {author} {\bibfnamefont
  {F.}~\bibnamefont {Vernon}},\ }\bibfield  {title} {\bibinfo {title} {Internal
  structure of the san jacinto fault zone in the trifurcation area southeast of
  anza, california, from data of dense seismic arrays},\ }\href@noop {}
  {\bibfield  {journal} {\bibinfo  {journal} {Geophys. J. Int}\ }\textbf
  {\bibinfo {volume} {213}},\ \bibinfo {pages} {98} (\bibinfo {year}
  {2018})}\BibitemShut {NoStop}%
\bibitem [{\citenamefont {Sharp}(1967)}]{sharp1967}%
  \BibitemOpen
  \bibfield  {author} {\bibinfo {author} {\bibfnamefont {R.~V.}\ \bibnamefont
  {Sharp}},\ }\bibfield  {title} {\bibinfo {title} {San jacinto fault zone in
  the peninsular ranges of southern california},\ }\href@noop {} {\bibfield
  {journal} {\bibinfo  {journal} {Geological Society of America Bulletin}\
  }\textbf {\bibinfo {volume} {78}},\ \bibinfo {pages} {705} (\bibinfo {year}
  {1967})}\BibitemShut {NoStop}%
\bibitem [{\citenamefont {Shapiro}\ and\ \citenamefont
  {Kneib}(1993)}]{shapiro1993}%
  \BibitemOpen
  \bibfield  {author} {\bibinfo {author} {\bibfnamefont {S.}~\bibnamefont
  {Shapiro}}\ and\ \bibinfo {author} {\bibfnamefont {G.}~\bibnamefont
  {Kneib}},\ }\bibfield  {title} {\bibinfo {title} {Seismic attenuation by
  scattering: theory and numerical results},\ }\href@noop {} {\bibfield
  {journal} {\bibinfo  {journal} {Geophys. J. Int}\ }\textbf {\bibinfo {volume}
  {114}},\ \bibinfo {pages} {373} (\bibinfo {year} {1993})}\BibitemShut
  {NoStop}%
\bibitem [{\citenamefont {Aki}(1969)}]{aki1969}%
  \BibitemOpen
  \bibfield  {author} {\bibinfo {author} {\bibfnamefont {K.}~\bibnamefont
  {Aki}},\ }\bibfield  {title} {\bibinfo {title} {Analysis of the seismic coda
  of local earthquakes as scattered waves},\ }\href@noop {} {\bibfield
  {journal} {\bibinfo  {journal} {J. Geophys. Res.}\ }\textbf {\bibinfo
  {volume} {74}},\ \bibinfo {pages} {615} (\bibinfo {year} {1969})}\BibitemShut
  {NoStop}%
\bibitem [{\citenamefont {Sato}\ \emph {et~al.}(2012)\citenamefont {Sato},
  \citenamefont {Fehler},\ and\ \citenamefont {Maeda}}]{sato2012}%
  \BibitemOpen
  \bibfield  {author} {\bibinfo {author} {\bibfnamefont {H.}~\bibnamefont
  {Sato}}, \bibinfo {author} {\bibfnamefont {M.~C.}\ \bibnamefont {Fehler}},\
  and\ \bibinfo {author} {\bibfnamefont {T.}~\bibnamefont {Maeda}},\
  }\href@noop {} {\emph {\bibinfo {title} {Seismic wave propagation and
  scattering in the heterogeneous earth}}},\ Vol.\ \bibinfo {volume} {496}\
  (\bibinfo  {publisher} {Springer},\ \bibinfo {year} {2012})\BibitemShut
  {NoStop}%
\bibitem [{\citenamefont {Kurzon}\ \emph {et~al.}(2014)\citenamefont {Kurzon},
  \citenamefont {Vernon}, \citenamefont {Ben-Zion},\ and\ \citenamefont
  {Atkinson}}]{kurzon2014}%
  \BibitemOpen
  \bibfield  {author} {\bibinfo {author} {\bibfnamefont {I.}~\bibnamefont
  {Kurzon}}, \bibinfo {author} {\bibfnamefont {F.~L.}\ \bibnamefont {Vernon}},
  \bibinfo {author} {\bibfnamefont {Y.}~\bibnamefont {Ben-Zion}},\ and\
  \bibinfo {author} {\bibfnamefont {G.}~\bibnamefont {Atkinson}},\ }\bibfield
  {title} {\bibinfo {title} {Ground motion prediction equations in the san
  jacinto fault zone: Significant effects of rupture directivity and fault zone
  amplification},\ }\href@noop {} {\bibfield  {journal} {\bibinfo  {journal}
  {Pure and App. Geophys.}\ }\textbf {\bibinfo {volume} {171}},\ \bibinfo
  {pages} {3045} (\bibinfo {year} {2014})}\BibitemShut {NoStop}%
\bibitem [{\citenamefont {Allam}\ \emph {et~al.}(2014)\citenamefont {Allam},
  \citenamefont {Ben-Zion}, \citenamefont {Kurzon},\ and\ \citenamefont
  {Vernon}}]{allam2014}%
  \BibitemOpen
  \bibfield  {author} {\bibinfo {author} {\bibfnamefont {A.}~\bibnamefont
  {Allam}}, \bibinfo {author} {\bibfnamefont {Y.}~\bibnamefont {Ben-Zion}},
  \bibinfo {author} {\bibfnamefont {I.}~\bibnamefont {Kurzon}},\ and\ \bibinfo
  {author} {\bibfnamefont {F.}~\bibnamefont {Vernon}},\ }\bibfield  {title}
  {\bibinfo {title} {Seismic velocity structure in the hot springs and
  trifurcation areas of the san jacinto fault zone, california, from
  double-difference tomography},\ }\href@noop {} {\bibfield  {journal}
  {\bibinfo  {journal} {Geophys. J. Int}\ }\textbf {\bibinfo {volume} {198}},\
  \bibinfo {pages} {978} (\bibinfo {year} {2014})}\BibitemShut {NoStop}%
\bibitem [{\citenamefont {Vernon}\ \emph {et~al.}(2014)\citenamefont {Vernon},
  \citenamefont {Ben-Zion},\ and\ \citenamefont {Hollis}}]{vernon2014}%
  \BibitemOpen
  \bibfield  {author} {\bibinfo {author} {\bibfnamefont {F.}~\bibnamefont
  {Vernon}}, \bibinfo {author} {\bibfnamefont {Y.}~\bibnamefont {Ben-Zion}},\
  and\ \bibinfo {author} {\bibfnamefont {D.}~\bibnamefont {Hollis}},\
  }\bibfield  {title} {\bibinfo {title} {Sage brush flats nodal experiment.
  international federation of digital seismograph networks. other/seismic
  network.},\ }\href {https://doi .org /10 .7914 /SN /ZG _2014} {\  (\bibinfo
  {year} {2014})}\BibitemShut {NoStop}%
\end{thebibliography}

\end{document}